\documentclass[11pt,prd,preprint,superscriptaddress]{revtex4}
\usepackage{revsymb}
\usepackage{graphicx}
\usepackage{bm}
\usepackage{amsmath}
\usepackage{amssymb}
\usepackage{graphicx}

\begin{document}

\title{MATTER POWER SPECTRUM FOR THE
GENERALIZED CHAPLYGIN GAS MODEL: THE NEWTONIAN APPROACH}

\author{J.C. Fabris\footnote{E-mail: fabris@cce.ufes.br. Present address:
Institut d'Astrophysique de Paris - IAP, Paris, France.}}
\affiliation{Universidade Federal do Esp\'{\i}rito Santo,
Departamento
de F\'{\i}sica\\
Av. Fernando Ferrari, 514, Campus de Goiabeiras, CEP 29075-910,
Vit\'oria, Esp\'{\i}rito Santo, Brazil}

\author{S.V.B.
Gon\c{c}alves\footnote{E-mail: sergio.vitorino@pq.cnpq.br. Present address: Laboratoire d'Annecy-le-Vieux de Physique Theorique - LAPTH,
Annecy-le-Vieux, France.}}
\affiliation{Universidade Federal do Esp\'{\i}rito Santo,
Departamento
de F\'{\i}sica\\
Av. Fernando Ferrari, 514, Campus de Goiabeiras, CEP 29075-910,
Vit\'oria, Esp\'{\i}rito Santo, Brazil}

\author{H.E.S.
Velten\footnote{E-mail: velten@cce.ufes.br}}
\affiliation{Universidade Federal do Esp\'{\i}rito Santo,
Departamento
de F\'{\i}sica\\
Av. Fernando Ferrari, 514, Campus de Goiabeiras, CEP 29075-910,
Vit\'oria, Esp\'{\i}rito Santo, Brazil}

\author{W. Zimdahl\footnote{E-mail: zimdahl@thp.uni-koeln.de}}
\affiliation{Universidade Federal do Esp\'{\i}rito Santo,
Departamento
de F\'{\i}sica\\
Av. Fernando Ferrari, 514, Campus de Goiabeiras, CEP 29075-910,
Vit\'oria, Esp\'{\i}rito Santo, Brazil}

\begin{abstract}
We model the cosmic medium as the mixture of a generalized
Chaplygin gas and a pressureless matter component. Within a
neo-Newtonian approach (in which, different from standard
Newtonian cosmology, the pressure enters the homogeneous and
isotropic background dynamics) we compute the matter power
spectrum. The 2dFGRS data are used to discriminate between unified
models of the dark sector (a purely baryonic matter component of
roughly 5 percent of the total energy content and roughly 95
percent  generalized Chaplygin gas) and different models, for
which there is separate dark matter, in addition to that accounted
for by the generalized Chaplygin gas. Leaving the corresponding
density parameters free, we find that the unified models are
strongly disfavored. On the other hand, using unified model
priors, the observational data are also well described, in
particular for small and large values of the generalized Chaplygin
gas parameter $\alpha$. The latter result is in agreement with a
recent, more qualitative but fully relativistic, perturbation
analysis in \cite{staro}.
\end{abstract}

\maketitle
\date{\today}

\section{Introduction}
\label{Introduction} The crossing of different cosmological
observations, in particular the anisotropy spectrum of the cosmic
microwave background radiation (CMBR), the luminosity distance of
supernovae of type Ia, gravitational lensing and baryonic acoustic
oscillations, indicates that around $95\%$ of the cosmic
substratum is not directly detectable through electromagnetic
emission \cite{han1,pad1,turner,sahni1}. As long as one accepts
General Relativity (GR) to be valid, the now widely accepted
conclusion is that most of the substance in the universe must be
of non-baryonic origin. This dynamically dominating non-baryonic
substratum is usually divided into two components: dark matter, a
pressureless, agglomerating component, being present in local
structures like galaxies and clusters of galaxies, and smoothly
distributed dark energy, an exotic fluid with negative pressure.
Dark matter is required in order to explain the observed anomalies
in the dynamics of galaxies and cluster of galaxies, as well as to
generate the large scale structures in the universe; dark energy
is required in order to account for the present stage of
accelerated expansion of the universe and for the position of the
first acoustic peak in the CMBR spectrum. The nature of these dark
components remains a mystery. For reviews on the subject see
\cite{bergstrom,silk}.
\par
Among the host of models that have been proposed for dark matter
and dark energy over the last years, there are unified models of
the dark sector according to which there is just one dark
component that simultaneously plays the role of dark matter and
dark energy. The most popular proposal along this line is the
Chaplygin gas, an exotic fluid with negative pressure that scales
as the inverse of the energy density \cite{moschella}. This
phenomenologically introduced equation of state can be given a
string theory based motivation \cite{jackiw}. It has also been
generalized in different phenomenological ways \cite{berto}.
Another example for a unification scenario for the dark sector is
a bulk viscous model of the cosmic substratum \cite{winfried}.
While the Chaplygin gas model (in its traditional and generalized
forms) has been very successful in explaining the supernovae type
Ia data \cite{colistete}, there are claims that it does not pass
the tests connected with structure formation because of predicted
but not observed strong oscillations of the matter power
spectrum \cite{ioav}. It should be mentioned, however, that
oscillations in the Chaplygin gas component do not necessarily
imply corresponding oscillations in the observed baryonic power
spectrum \cite{avelino}.
\par
The generalized Chaplygin gas is characterized by the equation of state
\begin{equation}
\label{eos}
p = - \frac{A}{\rho^\alpha} \quad.
\end{equation}
For $A>0$ the pressure $p$ is negative, hence it may induce an accelerated
expansion of the universe. The corresponding sound speed is
positive as long as $\alpha
> 0$.
Recently, a gauge-invariant analysis of the baryonic matter power
spectrum for generalized Chaplygin gas cosmologies was shown to be
compatible with the data for parameter values $\alpha \approx 0$
and $\alpha \geq 3$ \cite{staro}. This result seems to strengthen
the role of Chaplygin gas type models as competitive candidates
for the dark sector. The present work provides a further investigation along these
lines. While we shall rediscover the mentioned results of \cite{staro}, albeit in a different framework, we also extend the scope of the analysis in the following sense.
The authors of \cite{staro} have shown that Chaplygin gas cosmologies are consistent with the data from structure formation for certain parameter configurations.
Here we ask additionally, whether or not the data really favor generalized Chaplygin gases as unified models of the dark sector. By leaving the density parameters of the Chaplygin gas and the non-relativistic matter component, respectively, free, we allow for a matter fraction that can be different from the pure baryonic part. This is equivalent to hypothetically admit the existence of an additional dark matter component. In other words, we do not prescribe the unified model from the start.
Moreover, our study
is not restricted to the spatially flat case. The 2dFGRS data are
then used to test whether or not the unified model, requiring that
the matter component describes baryonic matter with a density
parameter of the order of 5 percent only, is favored.

Now, a precise
estimation of the cosmological parameters using the matter power
spectrum is very involved, since a detailed discussion of many
physical processes like free streaming of neutrinos, electron
diffusion, etc. is necessary. We shall avoid such more complex
analysis by using conveniently the BBKS transfer function
\cite{bbks,sola,saulo} to impose the initial conditions. We
believe that this type of analysis retains the essential features
of the process and leads to quantitatively relevant results.

Our study relies on a {\it neo-Newtonian} approach which
represents a major simplification of the problem. In some sense, the neo-Newtonian
equations can be seen as the introduction of a first order
relativistic correction to the usual Newtonian equations \cite{harrison}. The
neo-Newtonian equations for cosmology
\cite{mccrea,harrison,lima,reis} modify the  Newtonian
equations in a way that makes the pressure
dynamically relevant already for the homogeneous and isotropic
background.
This allows us to describe an accelerated expansion of the Universe as the consequence of a sufficiently large effective negative pressure in a Newtonian framework.
While the neo-Newtonian approach reproduces the GR background dynamics exactly, differences occur at the perturbative level.
However, the GR first-order perturbation dynamics and its neo-Newtonian counterpart coincide exactly in the case of a vanishing sound speed \cite{reis}. One may therefore expect that the neo-Newtonian perturbation dynamics reproduces the correct GR results on all perturbation scales at least for small values of the sound speed. For constant equations of state it has been demonstrated that the correct large-scale behavior in the synchronous gauge is reproduced \cite{lima}.
On small scales one expects the spatial pressure gradient term to be relevant and the difference to the GR dynamics should be of minor importance.
Since the observational data correspond to modes that
are well inside the Hubble radius, the use of a Newtonian type
approach seems therefore adequate.

On this basis our analysis extends previous neo-Newtonian studies
to the two-component case. One of the components is a generalized
Chaplygin gas, the other one represents pressureless matter. The
advantage of employing a neo-Newtonian approach is a gain in
simplicity and transparency. While in future work all results will
have to be confirmed within GR, we shall ensure already here that
in the region of overlap between GR and neo-Newtonian dynamics our
results coincide with the corresponding GR results. Our
neo-Newtonian approach reproduces the parameter estimations for
the unified dark matter/dark energy in \cite{staro} also
numerically. Backed up by this success of the neo-Newtonian
approach we then enlarge the scope of our analysis and test the
validity of the unified model itself by relaxing the unified model
priors used in \cite{staro}. Denoting the present value of the
Chaplygin gas density parameter by $\Omega_{c0}$, we admit the
total present matter density parameter $\Omega_{m0}$ to be the sum
of an additional dark matter component with density parameter
$\Omega_{dm0}$ and the baryon contribution $\Omega_{b0}$, i.e.,
$\Omega_{m0} = \Omega_{dm0} + \Omega_{b0}$. Leaving the density
parameters free, we investigate whether or not the unified model
with $\Omega_{c0} \approx 0.96$, $\Omega_{b0} \approx 0.04$ and
$\Omega_{dm0} \approx 0$ is favored by the large-scale structure
data. We mention that a similar investigation using supernova
type Ia data reveals that the unification scenario is the most
favored one \cite{colistete}.
\par
Our Chaplygin gas cosmology has four free parameters: the value of
$\alpha$, the present Chaplygin gas and dark matter density
parameters $\Omega_{c0}$ and $\Omega_{dm0}$, respectively, and the
present Chaplygin gas sound speed $v^2_0$. There are two main
observational sources concerning the matter power spectrum today:
the 2dFGRS and the SDSS data sets \cite{cole,teg}. For reasons to
be discussed later, we will mainly use the 2dFGRS data.  For $\Omega_{dm0} \approx 0$,
$\Omega_{b0} \approx 0.04$ and  $\Omega_{c0} \approx 0.96$, equivalent to
the unified model, we obtain a very good fit of the date where
very small or very large values of $\alpha$ are preferred. This
reproduces the GR results of \cite{staro} in a Newtonian context.
On the other hand, when $\Omega_{dm0}$, $\Omega_{c0}$ and $\alpha$
are left free, large values of $\Omega_{dm0}$ and small values of
$\Omega_{c0}$ are preferred, thus disfavoring the unification
model. The same result is obtained when all four parameters
($\Omega_{dm0}$, $\Omega_{c0}$, $\alpha$ and $\bar A$) are left
free. If the curvature is fixed to zero, as indicated by the WMAP
results \cite{wmap}, implying  $\Omega_{c0} \approx 1 - \Omega_{m0}$,
the predictions do not change substantially and a scenario with
almost no dark energy is again preferred. In all cases, including
those for which the unification scenario is imposed from the
beginning, the minimum values of the $\chi^2$ statistical
parameter are very similar. This does not seem to allow definite
predictions of the model. Any conclusion seems to depend on the
chosen priors. We compare our results with those obtained from the
$\Lambda$CDM model, for which the power spectrum test indicates
$\Omega_{dm0} \approx 0.25$.  However, almost no restrictions on the
value of the cosmological constant are obtained. In fact, the
matter power spectrum seems to be a good indicator for dark matter
but not for dark energy.

The paper is organized as follows. In section \ref{gcg} we recall
the generalized Chaplygin gas model and the basic equations of standard Newtonian cosmology.
In section \ref{neon} we introduce the neo-Newtonian
framework for the two-component model of a generalized Chaplygin gas and pressureless matter and
establish the perturbation equations for this system.
In section \ref{powersp} the power spectrum is determined, from which the
probability distribution functions for each parameter are obtained. Our
results are discussed in section \ref{powersp}.

\section{The generalized Chaplygin gas model}
\label{gcg}

The generalized Chaplygin gas is characterized by the equation of state  (\ref{eos}), implying a negative pressure and a positive sound speed as long as $A>0$ and $\alpha
> 0$.
The observational constraints
from supernova type Ia data indicate that negative
values for $\alpha$ are favored, but the dispersion is high enough
to allow for a large range of positive values for this parameter
\cite{colistete}. Negative values for $\alpha$ imply an imaginary
sound velocity, leading to small scale instabilities at the perturbative level.
Rigourously, the general situation is more
complex: such instabilities for fluids with negative pressure may
disappear if the hydrodynamical approach is replaced by a more
fundamental description using, e,g., scalar fields.
However, this is not true for the Chaplygin gas: even in a
fundamental approach, using for example the Born-Infeld action,
the sound speed may be negative if $\alpha < 0$. For this reason
we shall not allow $\alpha$ to be negative.
\par
The traditional Chaplygin gas model is characterized by $\alpha =
1$. It is a consequence of the Nambu-Goto action parametrized in
light-cone coordinates. Through some suitable transformations, the light-cone parametrized
Nambu-Goto action reduces to the action
of a Newtonian fluid that obeys the equation of state (\ref{eos})
with $\alpha = 1$ \cite{jackiw}. In this sense, it is somehow
natural to construct a cosmological Chaplygin gas scenario within a Newtonian framework.
To be precise, the
symmetries of the Lagrangian are broken when gravity is included.
But this drawback can not even be cured by using a relativistic
version: in order to preserve the symmetries of the original
Nambu-Goto action a full string model must be implemented. But
the Newtonian approach remains a reasonable approximation because of the mentioned relation
between the Chaplygin gas and the Nambu-Goto action.
\par
Now, establishing a Newtonian model for a universe in accelerated expansion seems to be impossible.
In  traditional Newtonian
cosmology the pressure does not play any role in an
isotropic and homogeneous universe: the universe evolves always with the scale factor
$a(t) \propto t^{2/3}$, implying a decelerated expansion. This coincides with the
relativistic cosmology for a pressureless fluid. The pressure
becomes relevant only at perturbative level: there the
nature of the fluid is essential for the evolution of
the density contrast. For the evolution of
density perturbations in Newtonian cosmology see ref. \cite{weinberg}. The specific
application to the Chaplygin gas model has been discussed in
\cite{patricia}. Let us sketch its main lines.
The Newtonian cosmology is defined through the continuity
equation, the Euler equation and the Poisson equation
\cite{weinberg}:
\begin{eqnarray}
\label{em1}
\frac{\partial\rho}{\partial t} + \nabla\cdot(\rho\vec v) &=& 0 \ , \\
\label{em2}
\frac{\partial\vec v}{\partial t} + \vec v\cdot\nabla\vec v &=& - \frac{\nabla p}{\rho} - \nabla\phi \ , \\
\label{em3}
\nabla^2\phi &=& 4\pi G\rho \  .
\end{eqnarray}
In equations (\ref{em1})-(\ref{em3}) the pressure appears only in
the form of a gradient. Hence, the pressure itself does not enter the dynamics of a spatially homogeneous background, i.e., the equations do not
depend on the nature of the fluid. However, at perturbative level the relevant equation is
\begin{equation}
\ddot\delta + 2\frac{\dot a}{a}\dot\delta + \biggr\{\frac{k^2 v_s^2}{a^2} - 4\pi G\rho\biggl\}\delta = 0 \  ,
\end{equation}
where $\delta = \frac{\delta\rho}{\rho}$ is the density
contrast, $\delta\rho$ being a first order
fluctuation around the background solution, $v_s^2 = \partial p/\partial\rho$ is the sound velocity and $k$ is the wavenumber of the perturbation. If we consider
a fluid whose equation of state is given by
$p = \kappa \rho^\varepsilon$,
the solution is
\begin{equation}
\delta = t^{-\frac{1}{6}}\biggr\{c_1J_{5/(6\nu)}\left(\frac{\Lambda t^{-\nu}}{\nu}\right)
+ c_2J_{-5/(6\nu)}\left(\frac{\Lambda t^{-\nu}}{\nu}\right)\biggl\} \ ,
\end{equation}
with $\nu = \varepsilon - \frac{4}{3}$.

For the Chaplygin gas model the perturbations initially grow as in
the matter dominated universe, later they decrease, finally
approaching zero, which is the value for the cosmological constant
model \cite{patricia}.

\section{Neo-newtonian approach}
\label{neon}

The drawback of standard Newtonian cosmology,
the absence of a pressure term in the background dynamics,
has been cured in a simple way \cite{lima}: in the
conservation equation (\ref{em1}) one takes into account the work done by the
pressure during the expansion of the universe. At the same time, the
equation for the gravitational potential must be modified in
order to render the equations compatible. This has been done in
references \cite{mccrea,harrison,lima}. The final equations are
\begin{eqnarray}
\frac{\partial\rho}{\partial t} + \nabla\cdot(\rho\vec v) + p\nabla\cdot\vec v &=& 0 \ ,\\
\frac{\partial\vec v}{\partial t} + \vec v\cdot\nabla\vec v &=& - \frac{\nabla p}{\rho + p} - \nabla\phi \ ,\\
\nabla^2\phi &=& 4\pi G(\rho + 3p) \ .
\end{eqnarray}
For the case of two non-interacting fluids with energy densities $\rho_c$ and $\rho_m$ and pressures
$p_c$ and $p_m = 0$, respectively, the equations are:
\begin{eqnarray}
\label{neo1}
\frac{\partial\rho_c}{\partial t} + \nabla\cdot(\rho_c\vec v_c) + p_c\nabla\cdot\vec v_c &=& 0 \ ,\\
\label{neo2}
\frac{\partial\vec v_c}{\partial t} + \vec v_c\cdot\nabla\vec v_c &=& - \frac{\nabla p_c}{\rho_c + p_c} - \nabla\phi \ ,\\
\label{neo3}
\frac{\partial\rho_m}{\partial t} + \nabla\cdot(\rho_m\vec u_m) &=& 0 \ ,\\
\label{neo4}
\frac{\partial\vec v_m}{\partial t} + \vec v_m\cdot\nabla\vec v_m &=& - \nabla\phi \ ,\\
\label{neo5} \nabla^2\phi &=& 4\pi G(\rho_m + \rho_c + 3p_c) \
.
\end{eqnarray}
The subscript $m$ stands for pressureless matter and the subscript
$c$ for the (generalized) Chaplygin gas component. Considering now
an isotropic and homogeneous universe with $\rho = \rho(t)$, $p =
p(t)$ and $\vec v = \frac{\dot a}{a}\vec r$, we find
\begin{eqnarray}
\biggr(\frac{\dot a}{a}\biggl)^2 + \frac{k}{a^2} &=& \frac{8\pi
G}{3}(\rho_m + \rho_c) \ , \\
\frac{\ddot a}{a} &=& - \frac{4\pi G}{3}(\rho_c + \rho_m + 3p_c)
\ .
\end{eqnarray}
These equations are identical to the corresponding equations for a
homogeneous and isotropic universe in GR. In a
sense, the neo-Newtonian formulation intends to reproduce the
equations of GR, but in a Newtonian conceptual
framework.

While there is a complete equivalence between the general relativistic
and the neo-Newtonian equations in the homogeneous and isotropic
background, this is no longer the case at the perturbative level.
As already mentioned, the GR first-order perturbation dynamics and its neo-Newtonian counterpart coincide exactly only in the case of a vanishing sound speed \cite{reis}. But also for small values of the sound speed the neo-Newtonian perturbation dynamics will very likely be a reasonable approximation.
Since the observational data correspond to modes that
are well inside the Hubble radius, the use of a Newtonian type
approach seems adequate, at least as a first and transparent step towards a full relativistic treatment.

Defining the fractional density contrasts
\begin{equation}
\delta_c = \frac{\delta\rho_c}{\rho_c} \quad \mathrm{and }\quad \delta_m =
\frac{\delta\rho_m}{\rho_m} \
\end{equation}
for the Chaplygin gas and matter components, respectively, the first-order perturbation equations for the system
(\ref{neo1})-(\ref{neo5}) are
\begin{eqnarray}
\ddot\delta_c &+& \biggr\{2\frac{\dot a}{a} -
\frac{\dot\omega_c}{1 + \omega_c} + 3\frac{\dot a}{a}(v_c^2 -
\omega_c)\biggl\}\dot\delta_c
+ \left\{3\biggr(\frac{\ddot a}{a} + \frac{\dot a^2}{a^2}\biggl)(v_c^2 - \omega_c) \qquad\qquad\right.\nonumber\\
&&\left. + 3\frac{\dot a}{a}\biggr[\dot
v^2_c - \dot\omega_c\frac{(1 + v_c^2)}{1 + \omega_c}\biggl] +
\frac{v_c^2\,k^2}{a^2} - 4\pi G\rho_c(1 + \omega_c)(1 +
3v_c^2)\right\}\delta_c
= 4\pi G\rho_m(1 +
\omega_c)\delta_m
\label{dddc}
\end{eqnarray}
and
\begin{equation}
\ddot\delta_m + 2\frac{\dot a}{a}\dot\delta_m - 4\pi
G\rho_m\delta_m = 4\pi G\rho_m(1 + 3v_c^2)\delta_c \ ,
\label{dddm}
\end{equation}
where $v_c^2 = \frac{\partial p_c}{\partial\rho_c}$ and $\omega_c
= \frac{p_c}{\rho_c}$. The quantity $k^{2}$ denotes the square of
the comoving wave vector. Dividing the equations (\ref{dddc}) and
(\ref{dddm}) by $H_0^2$ and redefining the time as
$t\,H_0\rightarrow t$, these equations become dimensionless. In
terms of the scale factor $a$ as dynamical variable, the system
(\ref{dddc})-(\ref{dddm}) takes the form
\begin{eqnarray}
\label{dddcbis} \delta''_c &+& \biggr\{\frac{2}{a} + g(a) -
\frac{\omega_c'(a)}{1 + \omega_c(a)} - 3\frac{1 +
\alpha}{a}\omega_c(a)\biggl\}\delta'_c
\nonumber\\
&&\quad - \biggr\{3\biggr[\frac{g(a)}{a} + \frac{1}{a^2}\biggl](1
+ \alpha)\omega_c(a) + \frac{3}{a}\biggr(\frac{1 + \alpha}{1 +
\omega_c(a)}\biggl)\omega'_c(a) +
\frac{\alpha\omega_c(a)\,k^2l_H^2}{a^2\,f(a)}
\nonumber\\
&&\quad + \frac{3}{2}\frac{\Omega_{c0}}{f(a)}h(a)[1 + \omega_c(a)][1 - 3\alpha\omega_c(a)]\biggl\}\delta_c
= \frac{3}{2}\frac{\Omega_{m0}}{a^3\,f(a)}[1 + \omega_c(a)]\delta_m \
\end{eqnarray}
and
\begin{equation}
\delta''_m + \biggr[\frac{2}{a} + g(a)\biggl]\delta'_m -
\frac{3}{2}\frac{\Omega_{m0}}{a^3\,f(a)}\delta_m =
\frac{3}{2}\frac{\Omega_{c0}}{f(a)}h(a)[1 -
3\alpha\omega_c(a)]\delta_c \ ,\label{dddma}
\end{equation}
where $l_H = c\,H_{0}^{-1 }$ is the present Hubble radius and
$c$ is the velocity of light.
The prime denotes a derivative with respect to $a$ and the definitions
\begin{eqnarray}
f(a) &=& \frac{\dot a^2}{H_{0}^{2}} = \left[\frac{\Omega_{m0} + \Omega_{c0}a^3\,h(a)}{a} + \Omega_{k0}\right] \ ,\\
g(a) &=& \frac{\ddot a}{\dot a^2} = - \frac{\Omega_{m0} + \Omega_{c0}[h(a) - 3\bar A\, h^{-\alpha}]a^3}{2a[\Omega_{m0} + \Omega_{c0}a^3h(a) + \Omega_{k0}a]} \ ,\\
h(a) &=& [\bar A + (1 - \bar A)a^{-3(1 + \alpha)}]^\frac{1}{1 + \alpha} \ , \\
\omega_c(a) &=& - \frac{\bar A}{h(a)^{1 + \alpha}} \  ,
\end{eqnarray}
with
\begin{equation}
\bar A = \frac{A}{\rho_{c0}^{1 + \alpha}} \ , \quad v_{s0}^2 = \alpha\bar A \
\end{equation}
have been used. Recall that $\Omega_{m0} = \Omega_{dm0} +
\Omega_{b0}$. For the unified model to be an adequate description
one expects $\Omega_{m0} \approx \Omega_{b0}$. In case the data
indicate a substantial fraction of $\Omega_{dm0}$, the unified
model will be disfavored.

\section{The power spectrum: comparing the
theory with observations}
\label{powersp}

The power spectrum is defined by
\begin{equation}
{\cal P} = \delta_k^2 \quad ,
\end{equation}
where $\delta_k$ is the Fourier transform of dimensionless density
contrast $\delta_m$. We will constrain the free parameters using the quantity
\begin{equation}
\chi^2 = \sum_i\biggr(\frac{{\cal P}_i^o - {\cal P}_i^t}{\sigma_i}\biggl)^2
\quad ,
\label{chi}
\end{equation}
where ${\cal P}_i^o$ is the observational value for the power
spectrum, ${\cal P}_i^t$ is the corresponding theoretical result and
$\sigma_i$ denotes the error bar. The index $i$ refers to a measurement
corresponding to given wavenumber. The quantity (\ref{chi}) qualifies
the fitting of the observational data for a given theoretical
model with specific values of the free parameters.
Hence, $\chi^2$ is a function of the free parameters
of the model. The probability distribution function is then
defined as
\begin{equation}
F(x_n) = F_0\,e^{-\chi^2(x_n)/2} \quad ,
\end{equation}
where the $x_n$ denote the ensemble of free parameters and $F_0$
is a normalization constant.
In order to obtain an estimation for a given parameter one has to
integrate (marginalize) over all the other ones. For a more
detailed description of this statistical analysis see reference
\cite{colistete}.
\par
The 2dFGRS \cite{cole} and the SDSS \cite{teg} are the main
surveys to obtain matter power spectrum data. The last one covers
a larger range of scales but the error bars are more narrow for
the former one. There are some discussions in the literature
concerning the relation between the different data \cite{colebis}.
In fact, the use of one or the other or the combination of both
may result in different parameter estimations. For our model,
however, the difference in using one or the other set of data is
not significant (we have verified this!). Hence, from now on we
focus on the 2dFGRS observational data for the power spectrum. We
use the data that are related with the linear approximation, that
is, those for which $k\,h^{-1} \leq 0.185 Mpc^{-1}$, where $h$ is
defined by $H_{0} \equiv 100\cdot h \mathrm{km/s\cdot Mpc}$. This
definition should not be confused with the preceding definition of
the function $h(a)$.
\par
To fix the initial conditions we use the following procedure. The
$\Lambda$CDM power spectrum is well fitted using the BBKS transfer
function \cite{bbks}. Then, employing the perturbed equations for the $\Lambda$CDM  model
and integrating back from today to a distant past, say $z = 1.000$, we
fix the shape of the transfer function at that moment. The
spectrum determined in this way is then used as the initial
condition for our Chaplygin gas model. This procedure is described in
more detail in references \cite{sola,saulo}.
\par
\begin{center}
\begin{figure}[!t]
\begin{minipage}[t]{0.225\linewidth}
\includegraphics[width=\linewidth]{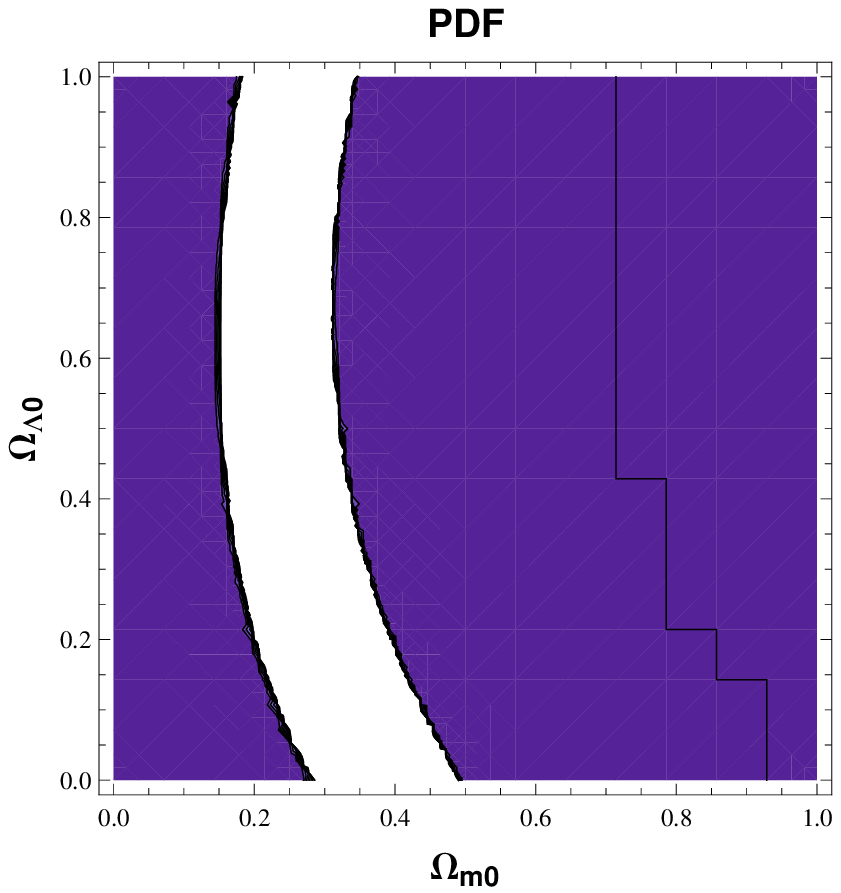}
\end{minipage} \hfill
\begin{minipage}[t]{0.225\linewidth}
\includegraphics[width=\linewidth]{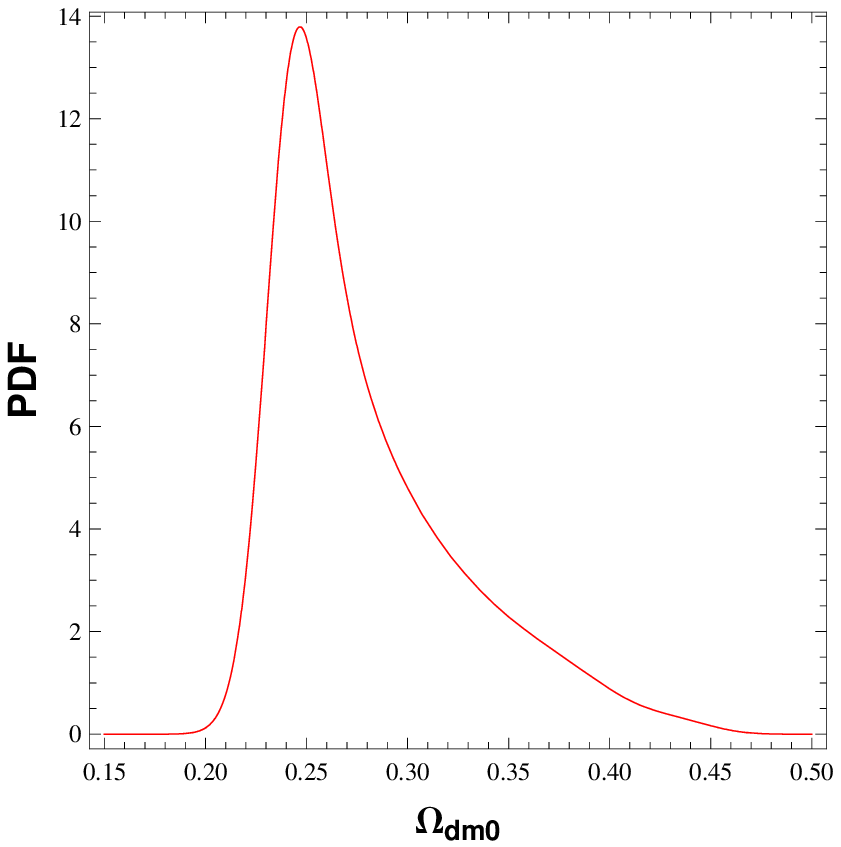}
\end{minipage} \hfill
\begin{minipage}[t]{0.225\linewidth}
\includegraphics[width=\linewidth]{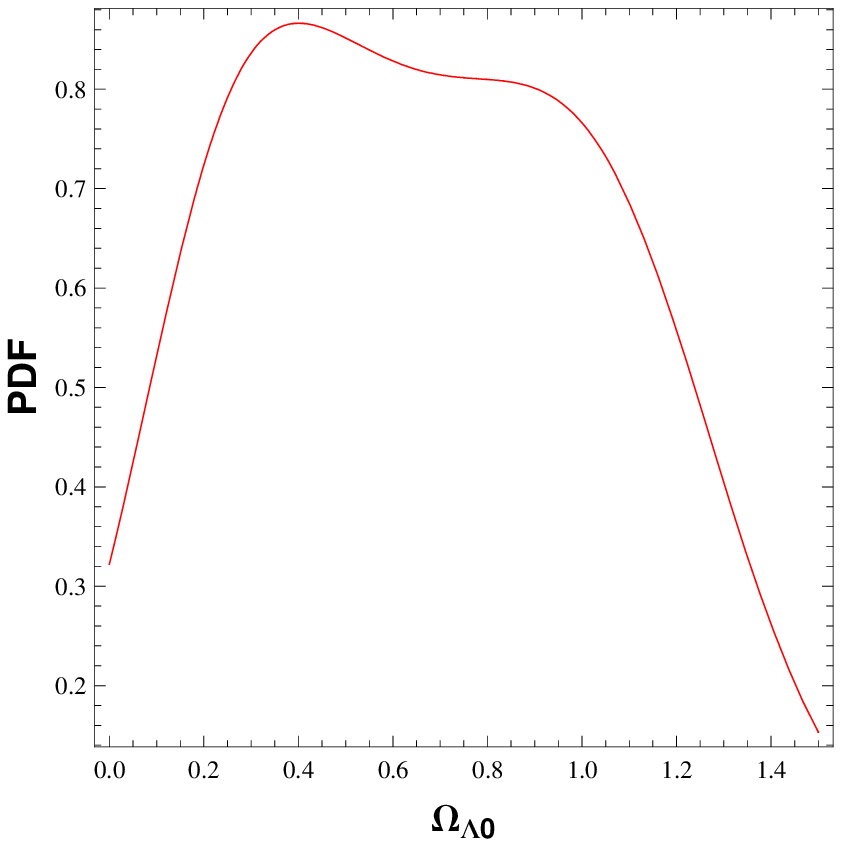}
\end{minipage} \hfill
\caption{{\protect\footnotesize The two-dimensional probability
distribution function (PDF) for $\Omega_{dm0}$ and
$\Omega_{\Lambda0}$ (left) and the corresponding one-dimensional
probability distribution functions for the non-flat $\Lambda$CDM
model. In the left panel: the darker  the color, the smaller the
probability.}} \label{LCDM}
\end{figure}
\end{center}

To ``gauge" our approach, let us first consider the $\Lambda$CDM
model. In the general (non-flat) case there are two parameters:
$\Omega_{dm0}$ and $\Omega_{\Lambda0}$. In figure \ref{LCDM}   we
show the two-dimensional probability distribution function (PDF)
as well the one-dimensional PDFs for the dark matter parameter
$\Omega_{dm0}$ and for the cosmological constant parameter
$\Omega_{\Lambda0}$, respectively. From the two dimensional
graphic it is clear that there is a large degeneracy for the
parameter $\Omega_{\Lambda0}$, while the region of allowable
values for $\Omega_{dm0}$ is quite narrow. The degeneracy for the
cosmological constant density is less visible in the
one-dimensional PDF graphic, but it is still considerable.
Incidentally, the minimum value for the $\chi^2$ parameter is
$0.3822$ for $\Omega_{dm0} = 0.2387$ and $\Omega_{\Lambda0} =
0.5937$, corresponding to an open universe.
\par
\begin{center}
\begin{figure}[!t]
\begin{minipage}[t]{0.225\linewidth}
\includegraphics[width=\linewidth]{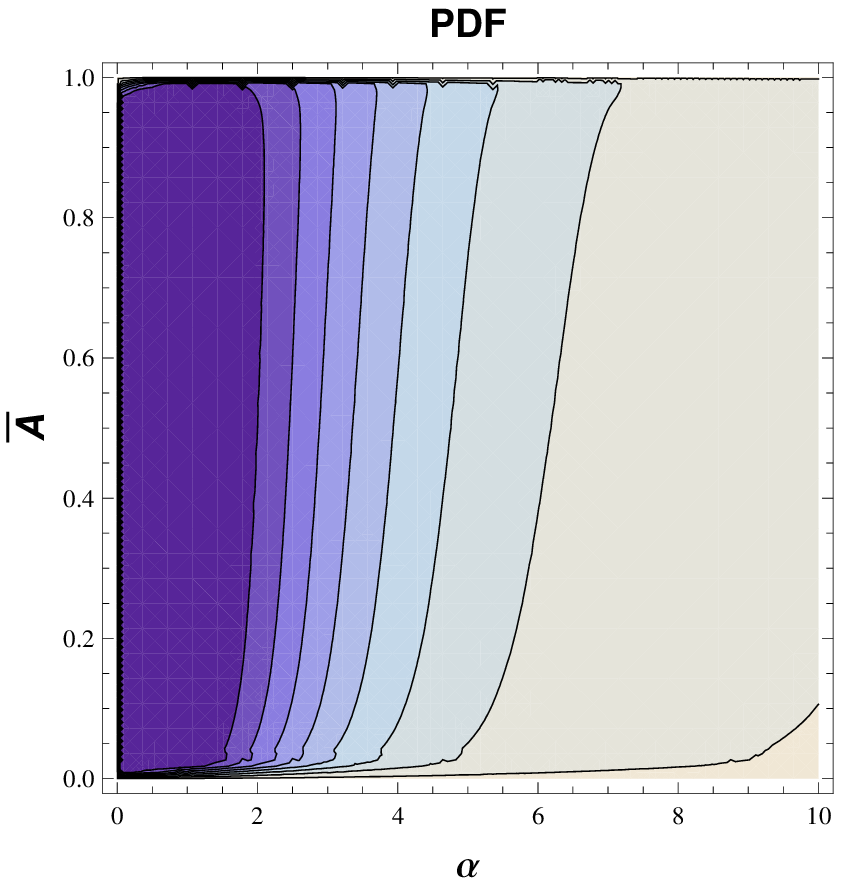}
\end{minipage} \hfill
\begin{minipage}[t]{0.225\linewidth}
\includegraphics[width=\linewidth]{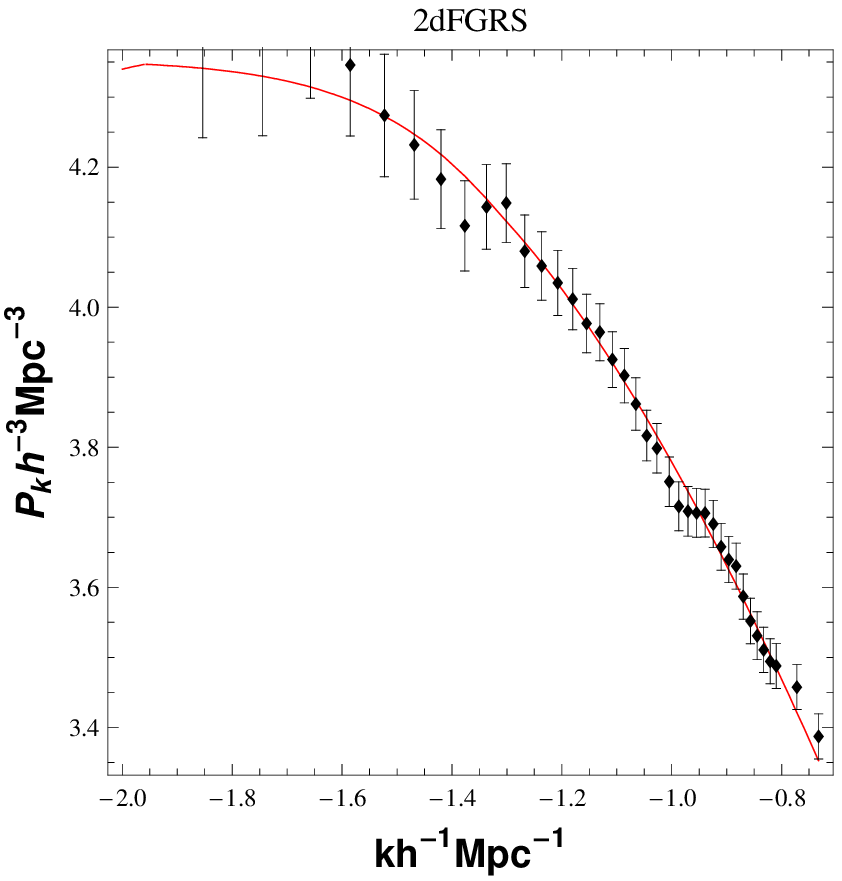}
\end{minipage} \hfill
\begin{minipage}[t]{0.225\linewidth}
\includegraphics[width=\linewidth]{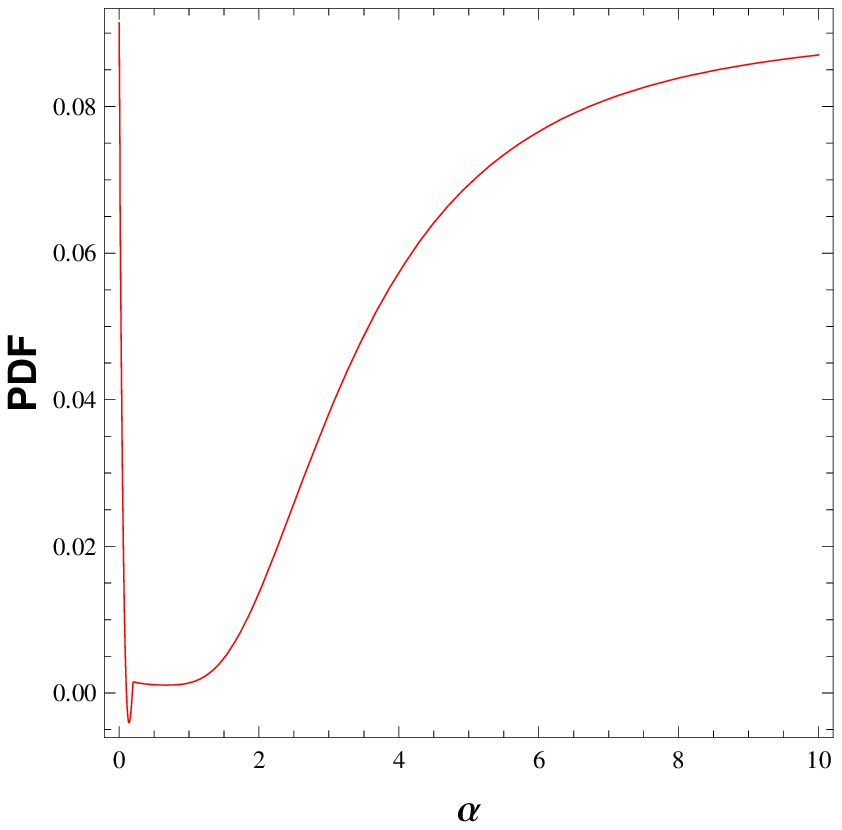}
\end{minipage} \hfill
\begin{minipage}[t]{0.225\linewidth}
\includegraphics[width=\linewidth]{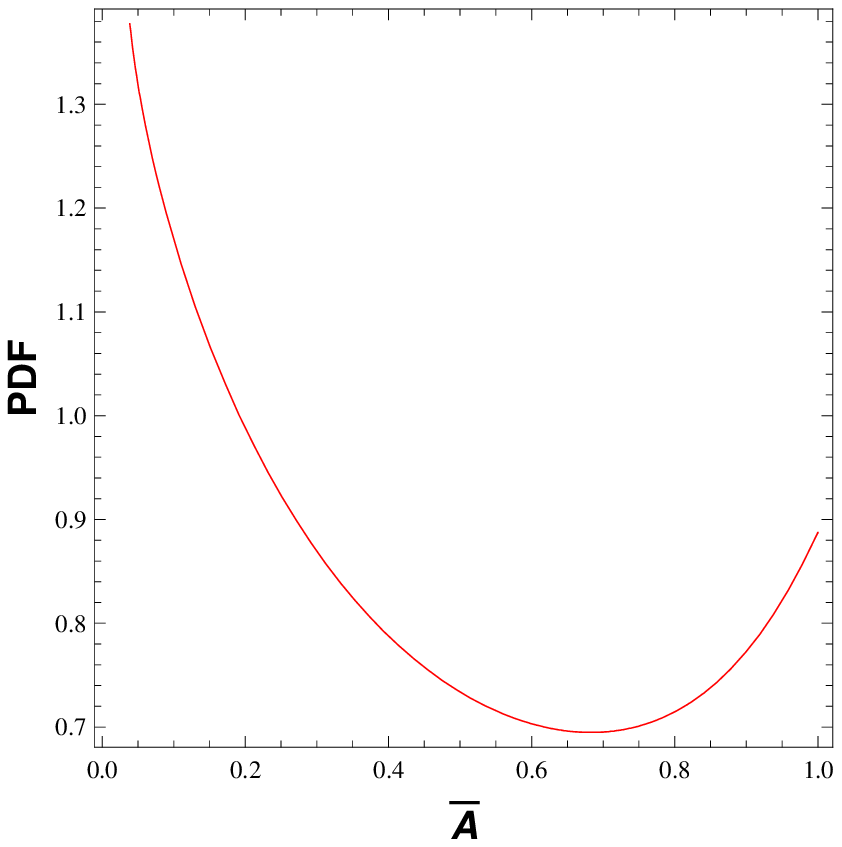}
\end{minipage} \hfill
\caption{{\protect\footnotesize The results for the flat case with
$\Omega_{b0} = 0.043$, $\Omega_{dm0} = 0$ and $\Omega_{c0} =
0.957$, corresponding to the unification scenario (case (i)). From
left to right: the two-dimensional PDF for $\alpha$ and $\bar A$
(with the same color convention as before), the best fitting curve
for the power spectrum, the one-dimensional PDFs for $\alpha$ and
$\bar A$. }} \label{UniPrior}
\end{figure}
\end{center}

The four free parameters to be constrained in our Chaplygin
gas model are $\Omega_{dm0}$, $\Omega_{c0}$, $\bar A$ and
$\alpha$. An analysis with four free parameters is computationally
hard. For this reason we shall start working with sets of three or
two free parameters, fixing the remaining one or two,
respectively. Only afterwards we consider the most general case in
which all parameters are left free. This strategy will allow us to
check the consistency of the final results. The baryonic component
$\Omega_{b0}$ is kept fixed in agreement with the nucleosynthesis
results. We use the value obtained by the recent five years WMAP
results, $\Omega_{b0} = 0.043$ (with $h = 0.72$). We will consider
the following cases: (i) a spatially flat universe with no
separate dark matter component, i.e. $\Omega_{dm0}= 0$,  a
baryonic component  given by $\Omega_{b0} = 0.043$ and a dark
sector component $\Omega_{c0} = 0.957$ - there are two free
parameters, $\alpha$ and $\bar A$; (ii) a flat universe with the
density parameters free, except for the condition $\Omega_{c0} = 1
- \Omega_{dm0} - \Omega_{b0}$, with the parameter $\bar A$ fixed
($\bar A = 0.15$ and $\bar A = 0.95$); (iii) a flat universe with
$\alpha$, $\bar A$ and one density parameter free; (iv) the
parameter $\bar A$ fixed (with values $0.15$ and $0.95$), while
$\alpha$ and the two density parameters are free; (v) all four
parameters free. Case (i) is the configuration studied in
reference \cite{staro}. Our results for the PDF essentially
confirm what was obtained in \cite{staro}: the one-dimensional
PDF, after marginalizing over $\bar A$, is higher near $\alpha =
0$ and for $\alpha
> 2$. For $\bar A$, the one-dimensional PDF, after marginalizing over $\alpha$, is initially high, then it decreases
until $\bar A \sim 0.7$, subsequently it is increasing again. This
behavior is shown in figure \ref{UniPrior}. Considering the
two-dimensional distribution, the minimum value for $\chi^2$ is
obtained for $\alpha = 3.57$ and $\bar A = 0$, with $\chi^2_{min}
= 0.378$, which is a value slightly better than that for the
$\Lambda CDM$ model, $\chi^2_{min} = 0.382$. However, the
superluminal sound speed renders this model unphysical. (For a
possible modification that could preserve causality, see
\cite{staro}).
\par
\begin{center}
\begin{figure}[!t]
\begin{minipage}[t]{0.225\linewidth}
\includegraphics[width=\linewidth]{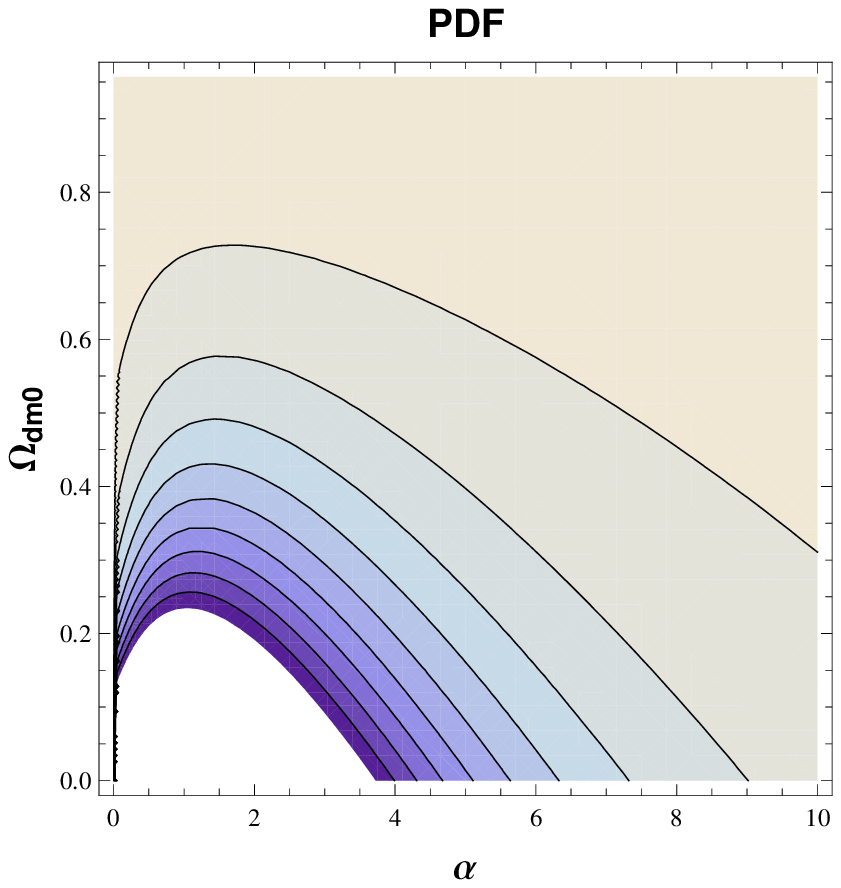}
\end{minipage} \hfill
\begin{minipage}[t]{0.225\linewidth}
\includegraphics[width=\linewidth]{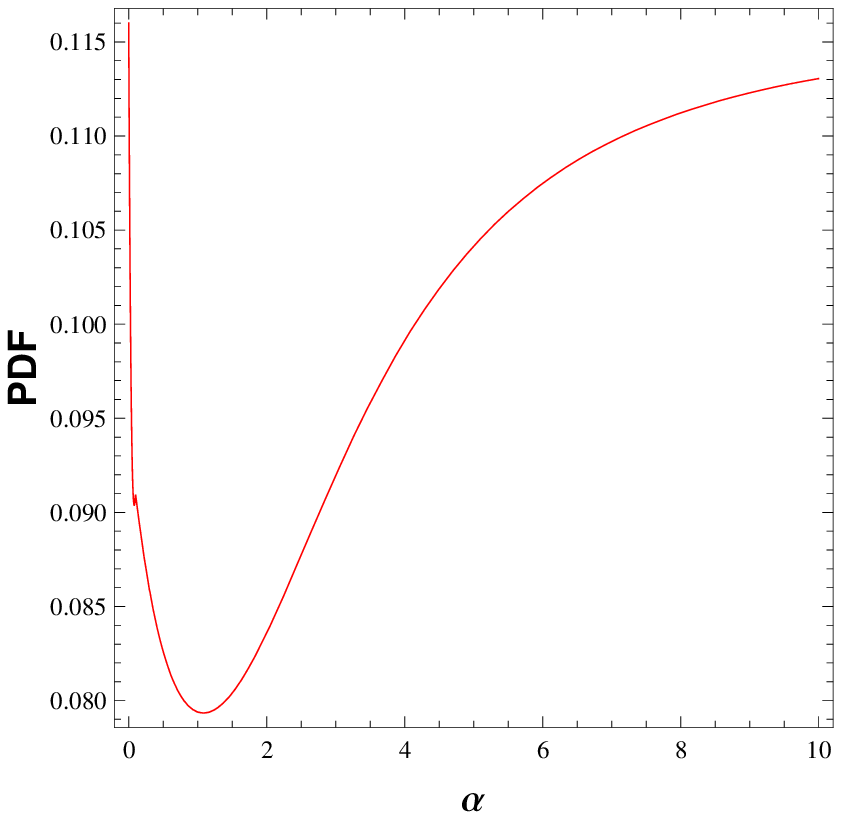}
\end{minipage} \hfill
\begin{minipage}[t]{0.225\linewidth}
\includegraphics[width=\linewidth]{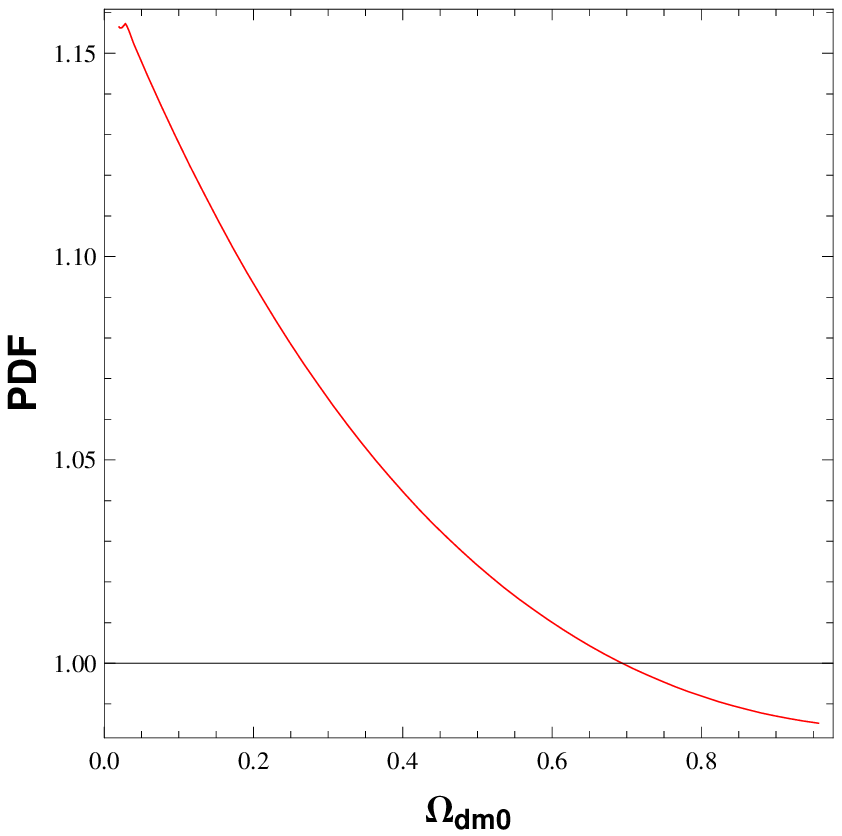}
\end{minipage} \hfill
\caption{{\protect\footnotesize The results for case (ii) with
$\Omega_{b0} = 0.043$, $\Omega_{dm0} = 1 - \Omega_{c0} -
\Omega_{b0}$. From left to right: the two-dimensional PDF for
$\alpha$ and $\Omega_{dm0}$, the one-dimensional PDFs for $\alpha$
and $\Omega_{dm0}$. }} \label{caseiia}
\end{figure}
\end{center}
\begin{center}
\begin{figure}[!t]
\begin{minipage}[t]{0.225\linewidth}
\includegraphics[width=\linewidth]{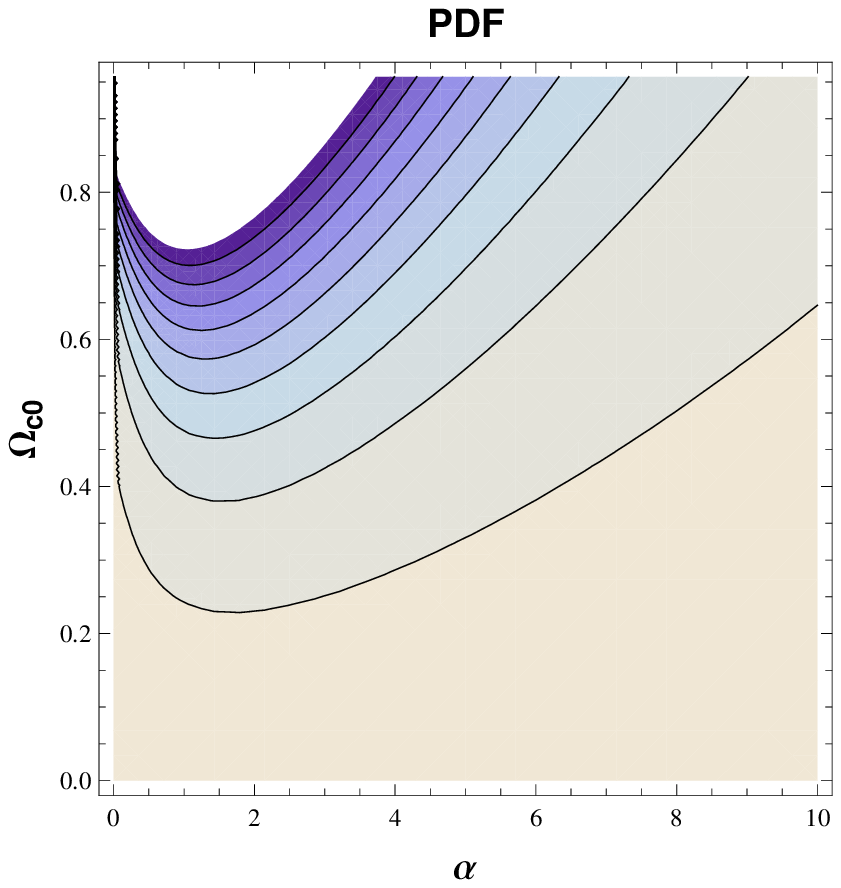}
\end{minipage} \hfill
\begin{minipage}[t]{0.225\linewidth}
\includegraphics[width=\linewidth]{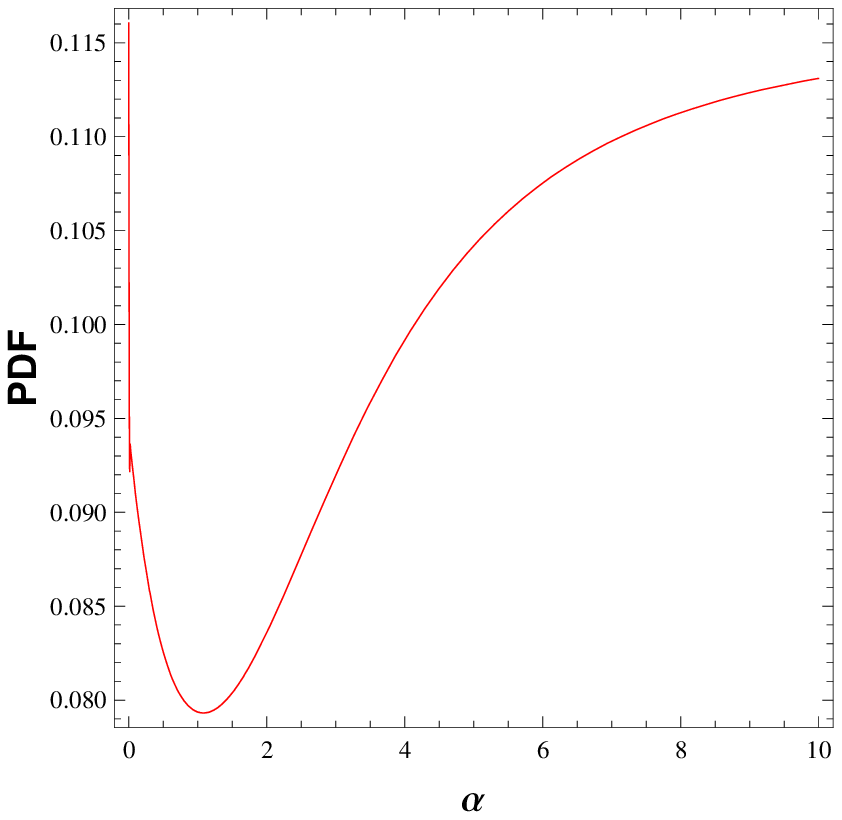}
\end{minipage} \hfill
\begin{minipage}[t]{0.225\linewidth}
\includegraphics[width=\linewidth]{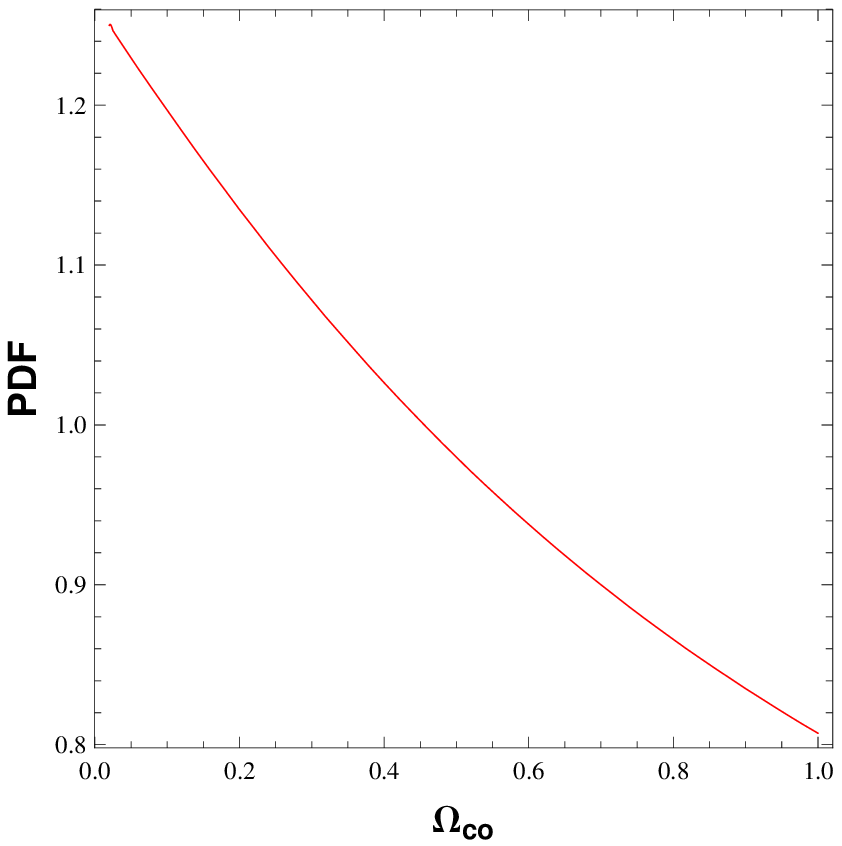}
\end{minipage} \hfill
\caption{{\protect\footnotesize The results for case (ii) with
$\Omega_{b0} = 0.043$, $\Omega_{c0} = 1 - \Omega_{md0} -
\Omega_{b0}$. From left to right: the two-dimensional PDF for
$\alpha$ and $\Omega_{c0}$, the one-dimensional PDFs for $\alpha$
and $\Omega_{c0}$. }} \label{caseiib}
\end{figure}
\end{center}

In case (ii) we relax the restriction that the pressureless
matter component is entirely given by baryons. It will turn out
that this leads to curious results. For vanishing spatial
curvature  $\Omega_{c0} = 1 - \Omega_{dm0} - \Omega_{b0}$ is
valid. As before, $\Omega_{b0}$ is fixed
 and we fix also $\bar A = 0.95$.
Now, varying $\Omega_{dm0}$, we span a two-dimensional PDF which
depends on $\alpha$ and $\Omega_{dm0}$. This two-dimensional PDF
and the corresponding one-dimensional PDFs for $\alpha$ and $\Omega_{dm0}$, respectively, are
shown in figure \ref{caseiia}. Again, values near $\alpha = 0$ and for $\alpha
> 2$ are favored. On the other hand,  the PDF for $\Omega_{dm0}$
decreases as $\Omega_{dm0}$  increases. This seems to favor the
unification scenario which requires a small $\Omega_{dm0}$.
However, if we vary $\Omega_{c0}$  instead of $\Omega_{dm0}$, we
find that the PDF for $\Omega_{c0}$ also decreases as
$\Omega_{c0}$ increases as is shown in figure \ref{caseiib}. This
seems to lead to the opposite conclusion than in the previous
case: now the unified scenario which requires a large
$\Omega_{c0}$ seems to be disfavored. Such a contradiction seems
to be an artifact of the marginalization procedure as can be seen
in the corresponding two-dimensional PDFs in figures \ref{caseiia}
and \ref{caseiib}: in the first case the probabilities are  high
near $\alpha = 0$ and for low values of $\Omega_{dm0}$, but at the
same time the minimum value for $\chi^2$ is obtained for $\alpha =
0$ and $\Omega_{dm0} = 1$; on the other hand, in the second case,
the probabilities are high near $\alpha = 0$ and $\Omega_{c0} =
0$, which are also the values for which the minimum of $\chi^2$ is
obtained. We conclude that under the given conditions the
unification scenario is disfavored. The same results are obtained
for $\bar A = 0.15$. In all these cases the minimum value for
$\chi^2$ is essentially the same as before.
\par
\begin{center}
\begin{figure}[!t]
\begin{minipage}[t]{0.225\linewidth}
\includegraphics[width=\linewidth]{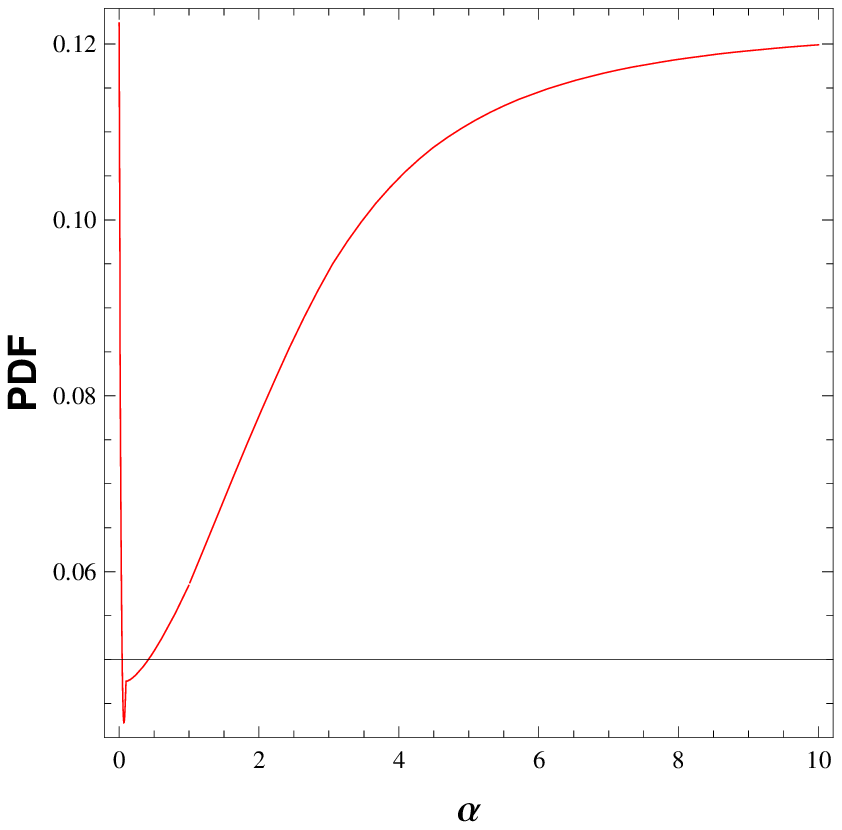}
\end{minipage} \hfill
\begin{minipage}[t]{0.225\linewidth}
\includegraphics[width=\linewidth]{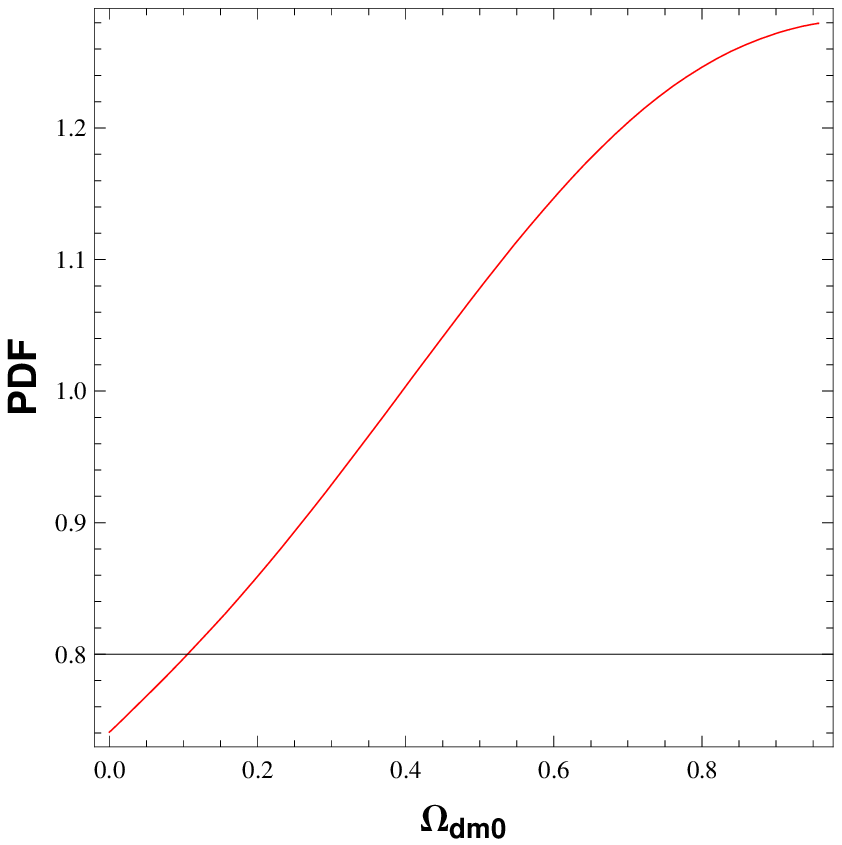}
\end{minipage} \hfill
\begin{minipage}[t]{0.225\linewidth}
\includegraphics[width=\linewidth]{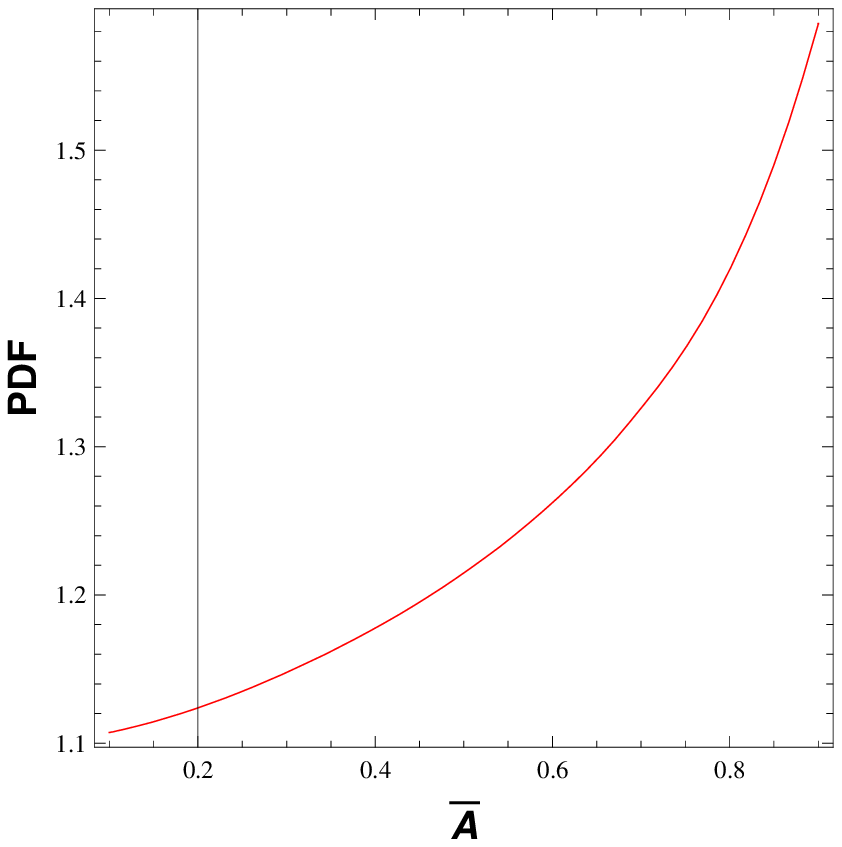}
\end{minipage} \hfill
\caption{{\protect\footnotesize The results for case (iii) with
$\Omega_{dm0} = 1 - \Omega_{c0} - \Omega_{b0}$. From left to
right: the one-dimensional PDFs for $\alpha$, $\Omega_{dm0}$ and
$\bar A$. }} \label{caseiiia}
\end{figure}
\end{center}
\begin{center}
\begin{figure}[!t]
\begin{minipage}[t]{0.225\linewidth}
\includegraphics[width=\linewidth]{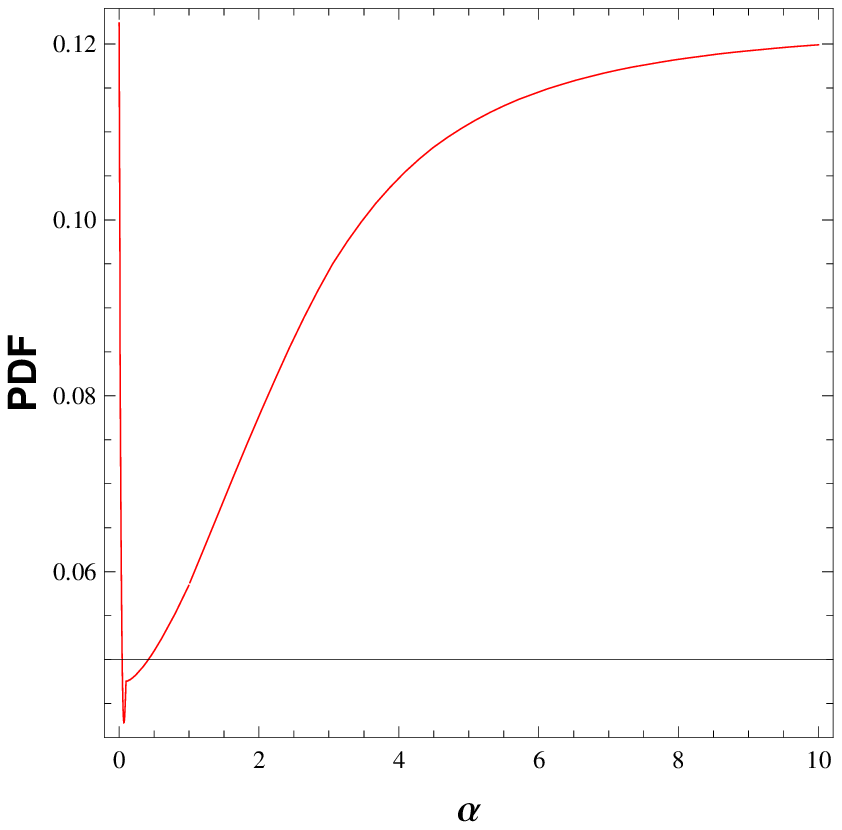}
\end{minipage} \hfill
\begin{minipage}[t]{0.225\linewidth}
\includegraphics[width=\linewidth]{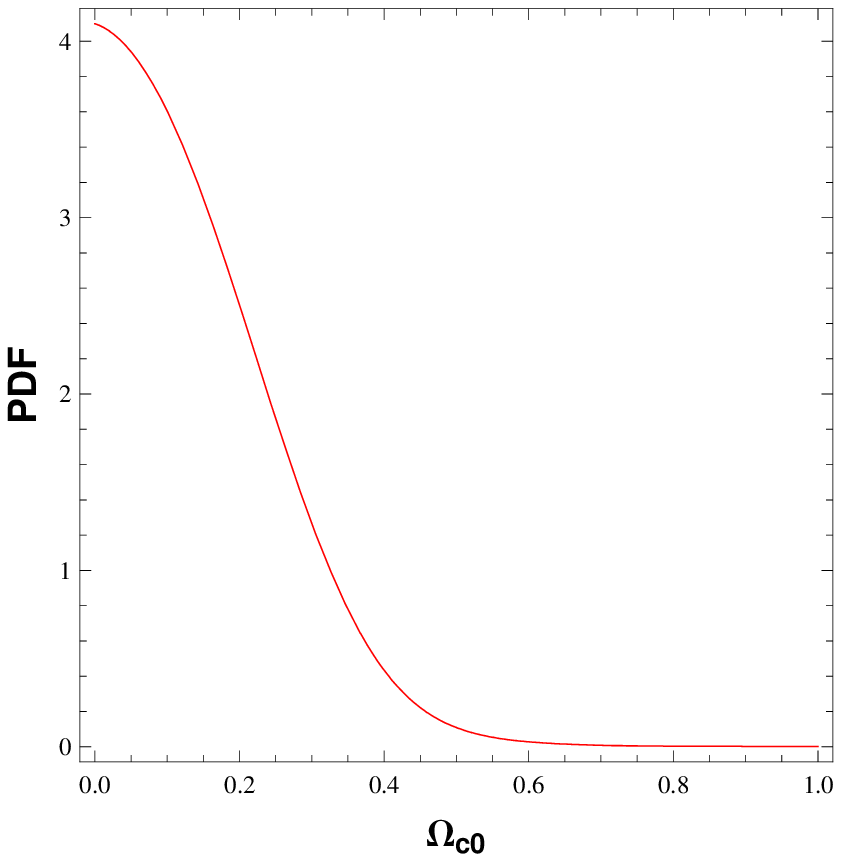}
\end{minipage} \hfill
\begin{minipage}[t]{0.225\linewidth}
\includegraphics[width=\linewidth]{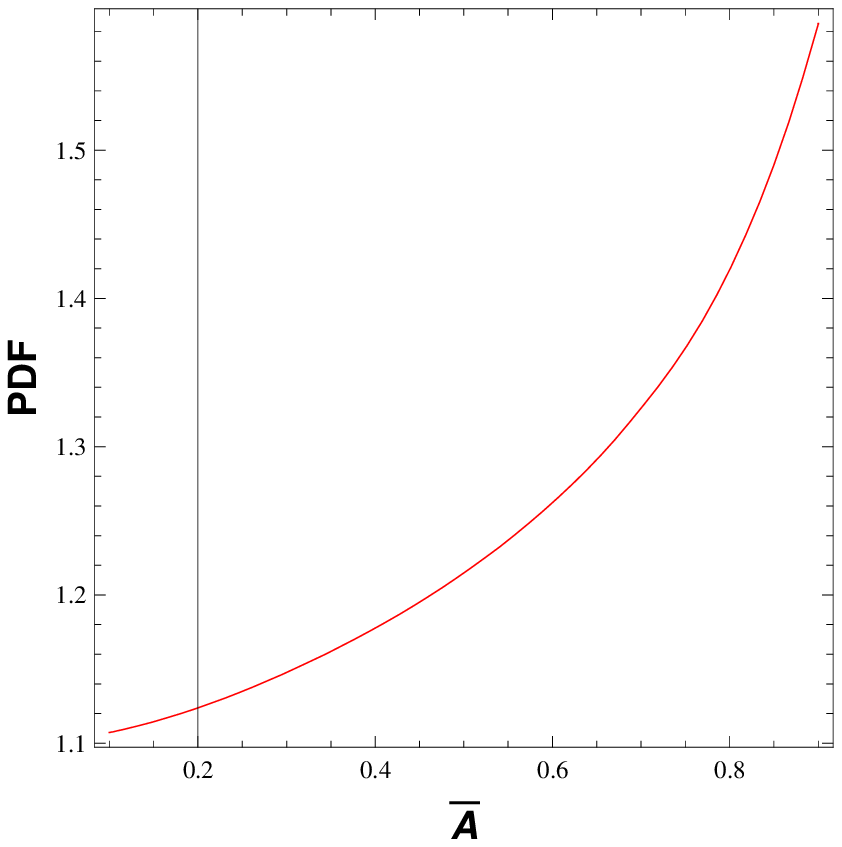}
\end{minipage} \hfill
\caption{{\protect\footnotesize The results for case (iii) with
$\Omega_{c0} = 1 - \Omega_{dm0} - \Omega_{b0}$. From left to
right: the one-dimensional PDFs for $\alpha$, $\Omega_{c0}$ and
$\bar A$. }} \label{caseiiib}
\end{figure}
\end{center}

To test the previous result, we construct again a
three-dimensional parameter space with zero spatial curvature but
leaving $\alpha$ and $\bar A$ free (case (iii)). Again, the PDF
for $\alpha$ is initially decreasing but increasing later for
$\alpha > 2$, while the PDF for $\Omega_{md0}$ is an increasing
function of $\Omega_{md0}$, as well as the PDF for $\bar A$
increases with $\bar A$. This result is shown in figure
\ref{caseiiia}. If we now vary $\Omega_{c0}$, its PDF is a
decreasing function of $\Omega_{c0}$ as shown in figure
\ref{caseiiib}.
\par
\begin{center}
\begin{figure}[!t]
\begin{minipage}[t]{0.225\linewidth}
\includegraphics[width=\linewidth]{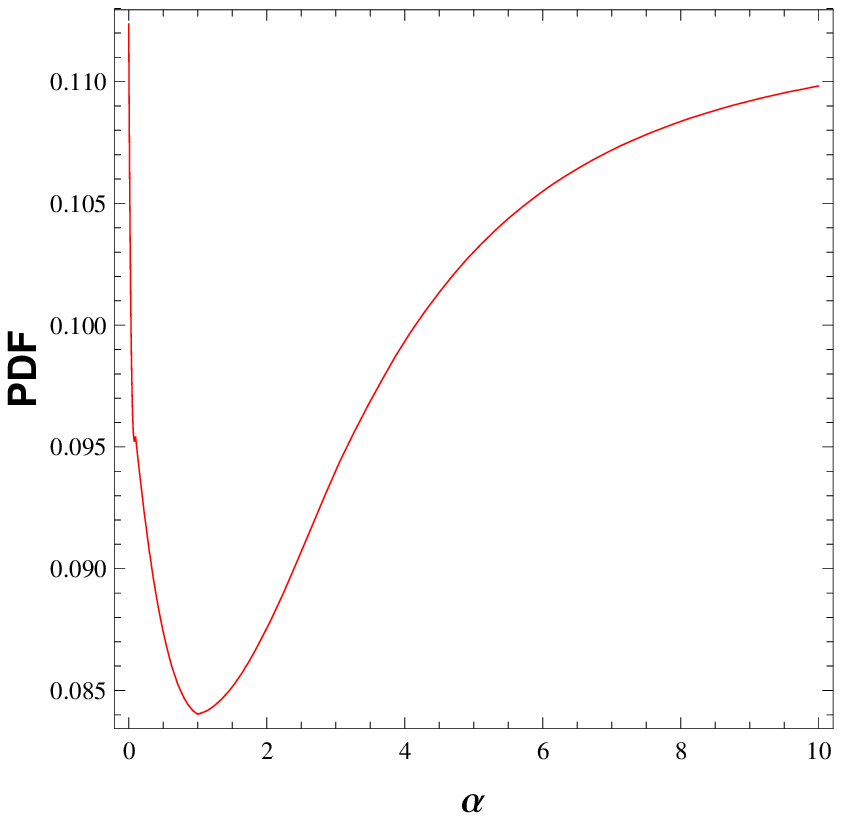}
\end{minipage} \hfill
\begin{minipage}[t]{0.225\linewidth}
\includegraphics[width=\linewidth]{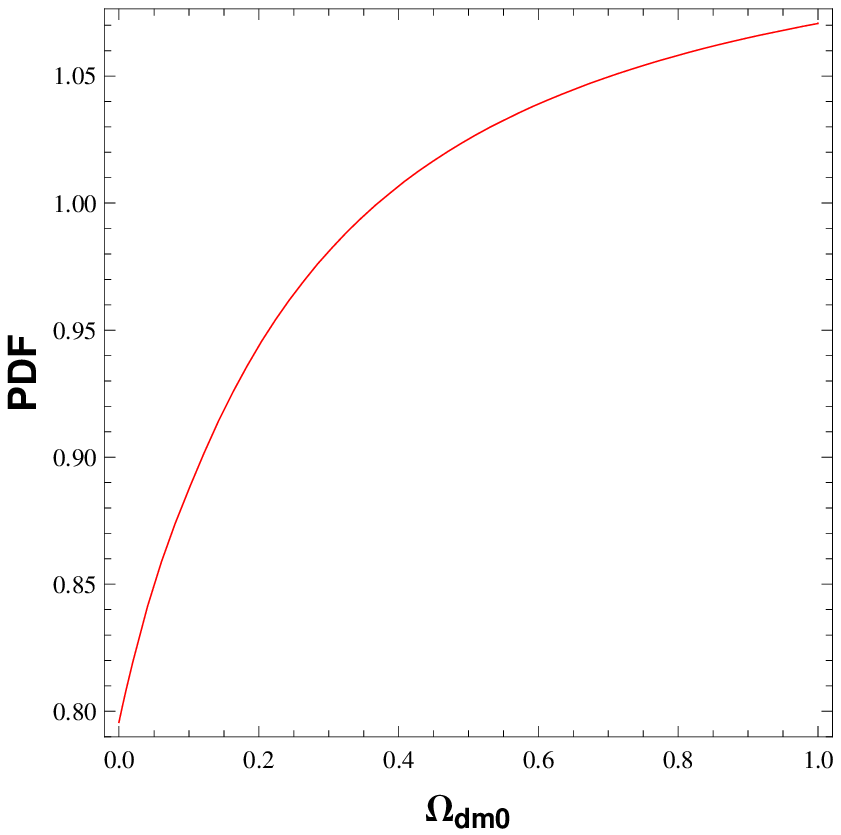}
\end{minipage} \hfill
\begin{minipage}[t]{0.225\linewidth}
\includegraphics[width=\linewidth]{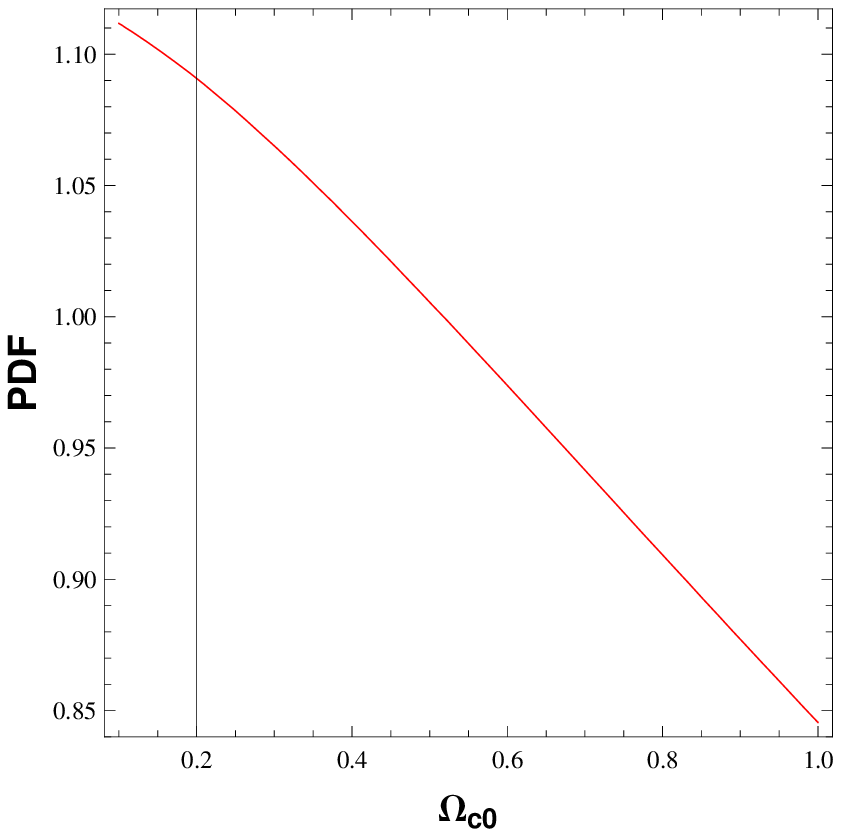}
\end{minipage} \hfill
\caption{{\protect\footnotesize The results for case (iv). From
left to right: the one-dimensional PDFs for $\alpha$,
$\Omega_{dm0}$ and $\Omega_{c0}$. }} \label{caseiv}
\end{figure}
\end{center}

The curious fact that the {\it anti-unification} scenario seems to
be favored is confirmed by another three-dimensional parameter
space study for which $\alpha$, $\Omega_{dm0}$ and $\Omega_{c0}$
are left free, while $\bar A = 0.95$ (case iv). Notice that the
curvature now is arbitrary. As can be seen in figure \ref{caseiv},
the PDF for $\alpha$ shows the same behavior as in the previous
cases, while the PDF for $\Omega_{dm0}$ now increases with
$\Omega_{dm0}$. On the other hand, the PDF for $\Omega_{c0}$
decreases with $\Omega_{c0}$. The {\it anti-unification} scenario
is clearly favored in this case, and there is no contradiction as
for configuration (ii). As in the previous cases, the minimum
value for $\chi^2$ is around $0.378$.
\par
\begin{center}
\begin{figure}[!t]
\begin{minipage}[t]{0.225\linewidth}
\includegraphics[width=\linewidth]{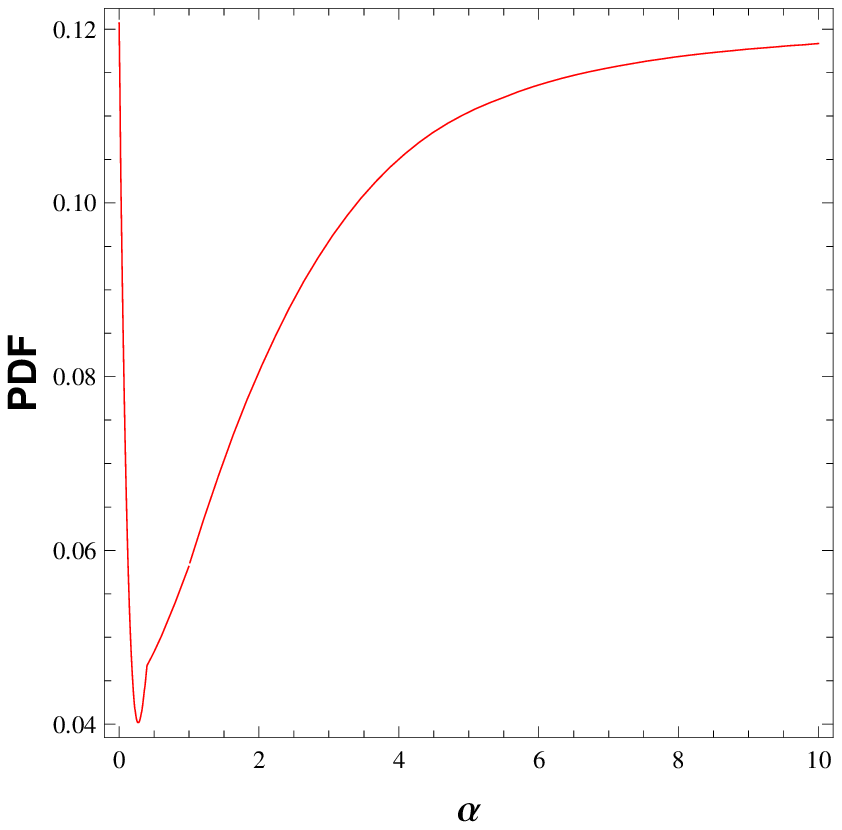}
\end{minipage} \hfill
\begin{minipage}[t]{0.225\linewidth}
\includegraphics[width=\linewidth]{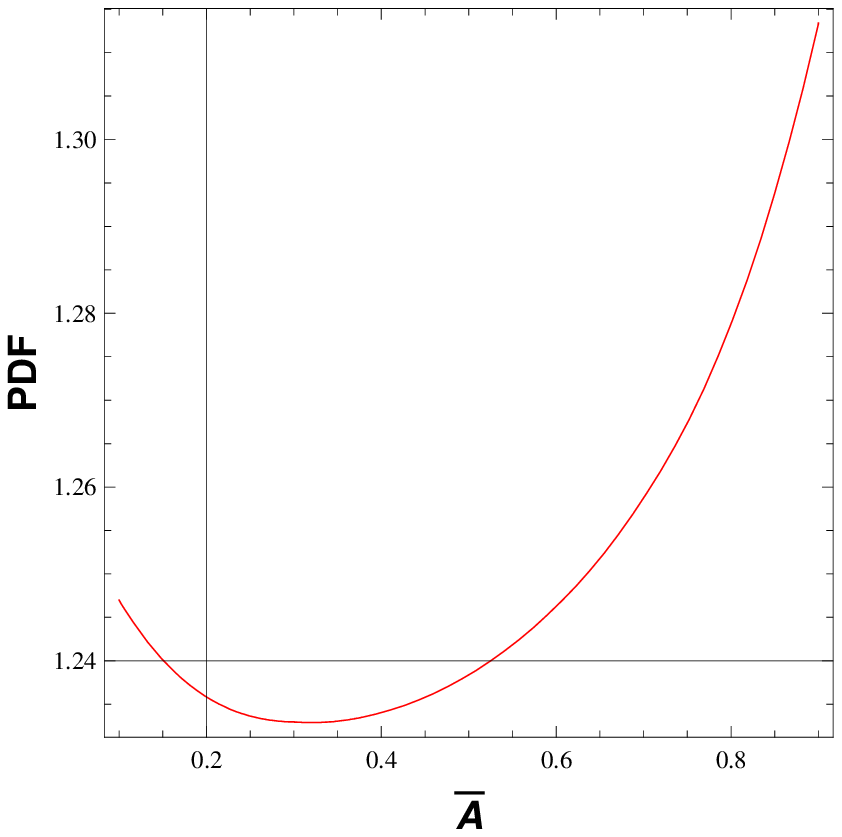}
\end{minipage} \hfill
\begin{minipage}[t]{0.225\linewidth}
\includegraphics[width=\linewidth]{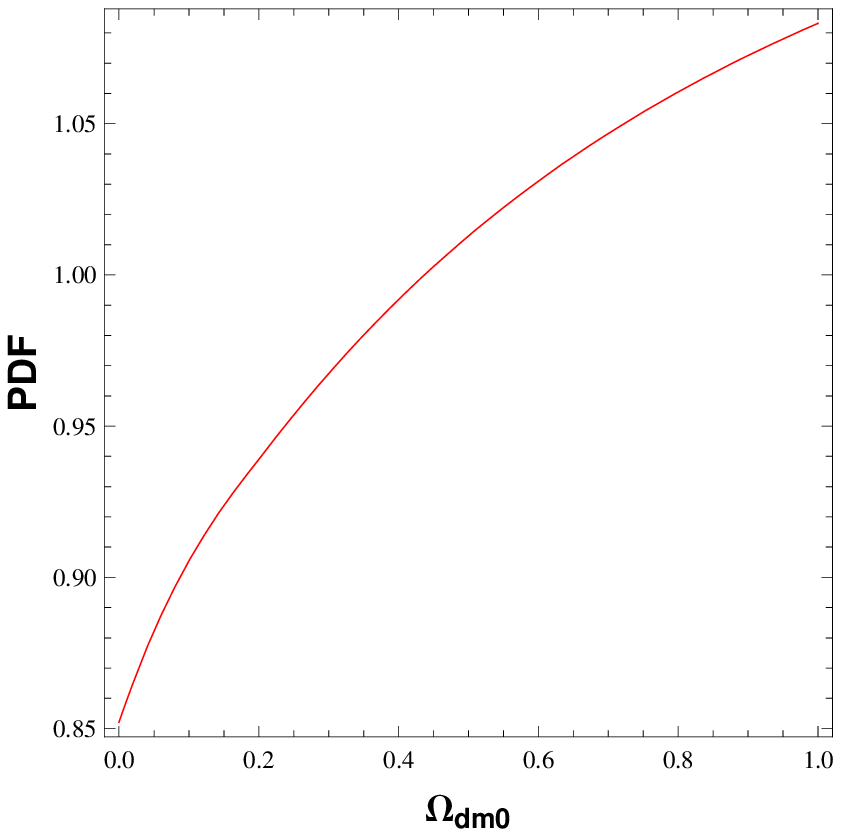}
\end{minipage} \hfill
\begin{minipage}[t]{0.225\linewidth}
\includegraphics[width=\linewidth]{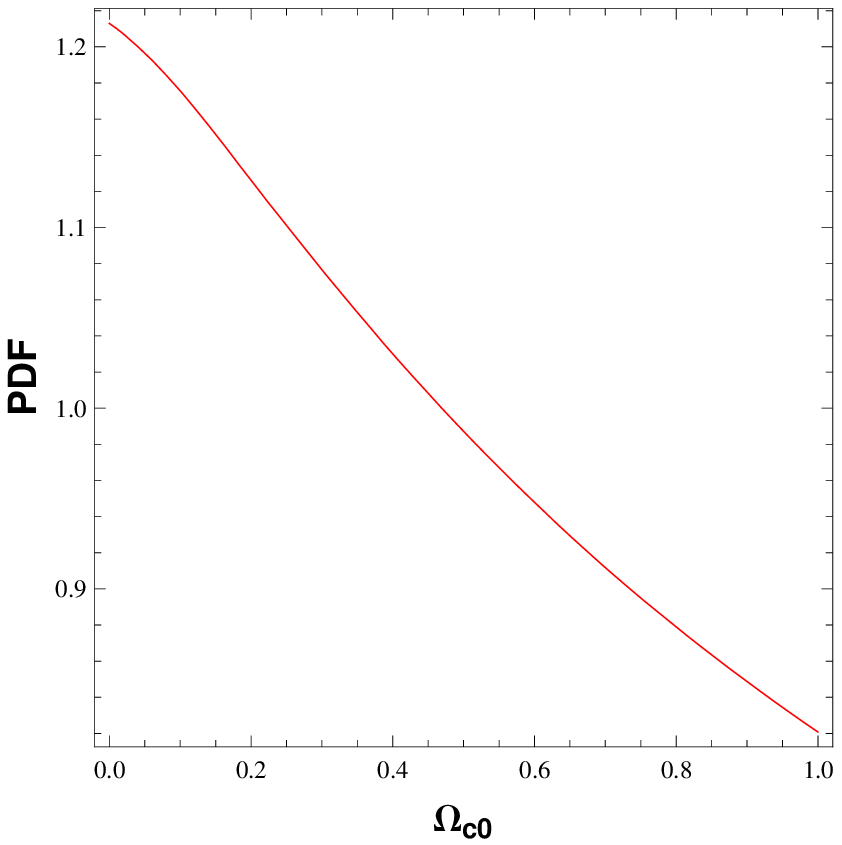}
\end{minipage} \hfill
\caption{{\protect\footnotesize The results for the general case
with four free parameters (case (v)). From left to right: the
one-dimensional PDFs  for $\alpha$, $\bar A$, $\Omega_{c0}$ and
$\Omega_{dm0}$.}} \label{4p}
\end{figure}
\end{center}

Varying all four parameters, all the preceding results are
confirmed. The one-dimensional PDFs for $\alpha$, $\bar A$,
$\Omega_{dm0}$ and $\Omega_{c0}$ are displayed in figure \ref{4p}.
It can be seen that the preferred values are either $\alpha \ll 1$
or $\alpha \geq 2$, while the probability is higher for large
values of $\Omega_{dm0}$ and small values of $\Omega_{c0}$.
\par
\begin{center}
\begin{figure}[!t]
\begin{minipage}[t]{0.225\linewidth}
\includegraphics[width=\linewidth]{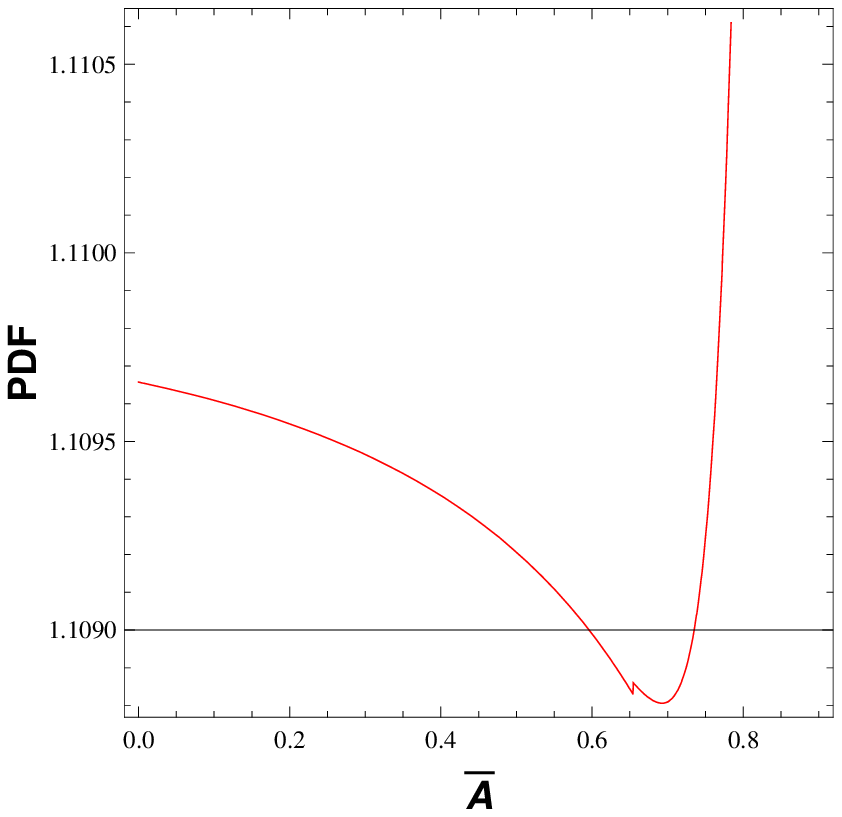}
\end{minipage} \hfill
\begin{minipage}[t]{0.225\linewidth}
\includegraphics[width=\linewidth]{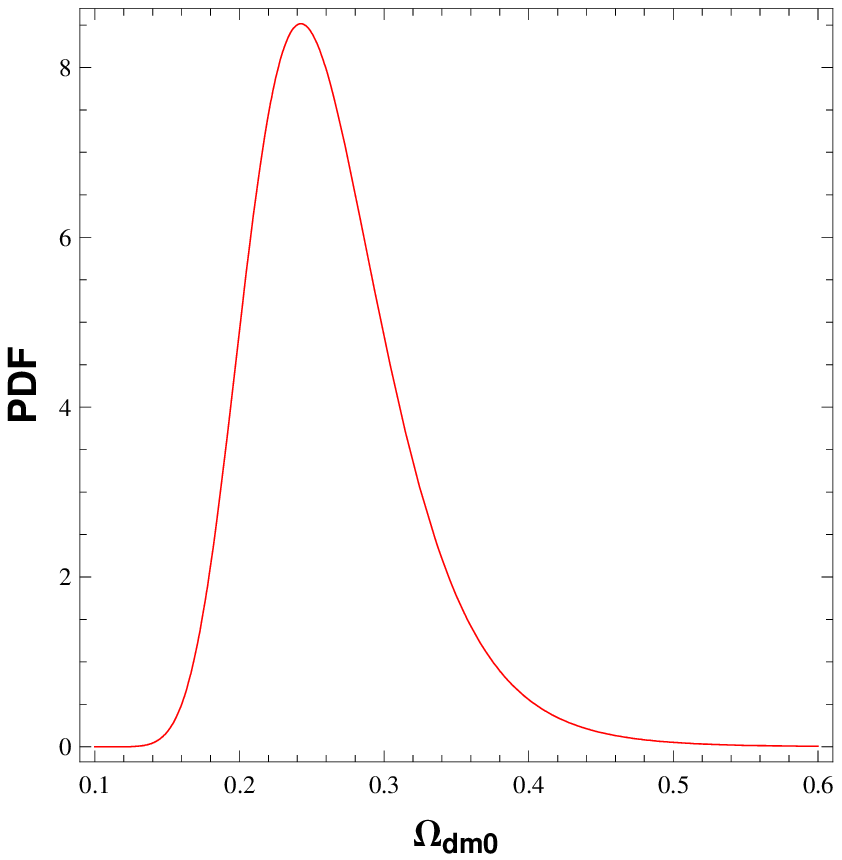}
\end{minipage} \hfill
\begin{minipage}[t]{0.225\linewidth}
\includegraphics[width=\linewidth]{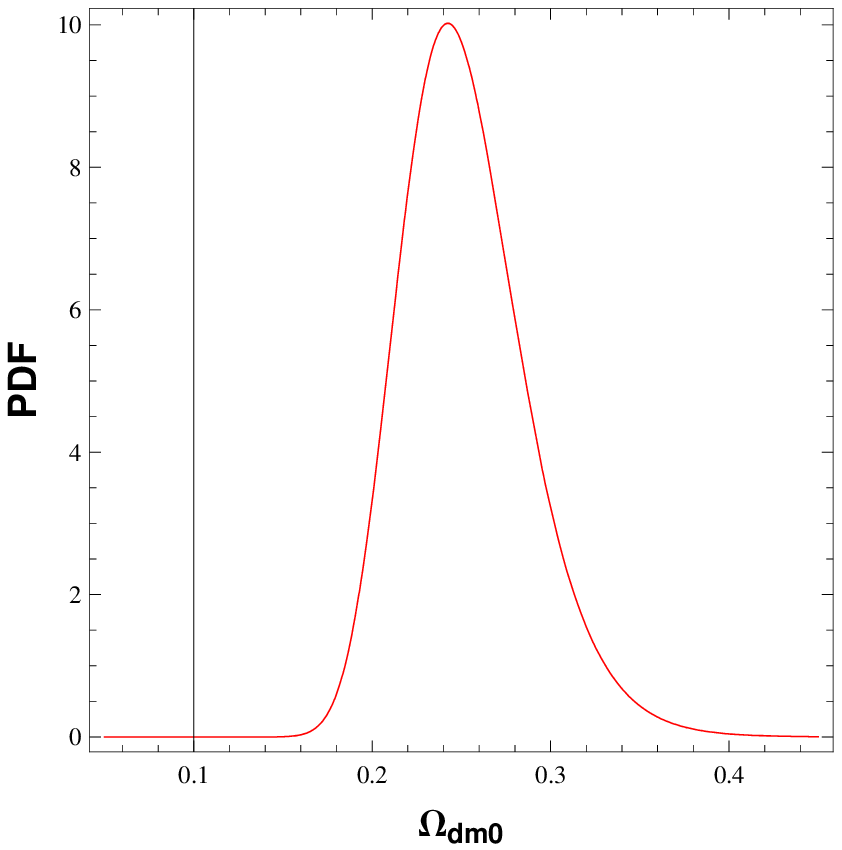}
\end{minipage} \hfill
\caption{{\protect\footnotesize The results for the flat case with $\alpha = 0$.
From left to right: the one-dimensional
PDFs  for $\bar A$, for
$\Omega_{dm0}$ with $\bar A \neq 1$ and for
$\Omega_{dm0}$ with $\bar A = 1$.}}
\label{alpha0}
\end{figure}
\end{center}

Finally, let us consider the particular case $\alpha = 0$. In
this situation, the neo-Newtonian perturbation dynamics exactly
coincides
 with that of GR.  The
$\Lambda$CDM model is recovered for $\bar A = 1$. The results for
$\alpha = 0$ and $\bar A \neq 1$ as well as for $\alpha = 0$ and
$\bar A = 1$, both for a flat universe, are displayed in figure
\ref{alpha0}: the predictions for $\Omega_{dm0}$ are essentially
the same as for the $\Lambda$CDM model. For $\bar A \neq 1$,
the probability for $\bar A$ grows as $\bar A$  approaches
unity. This shows that the method employed is consistent.
\section{Analysis of the results and conclusions}
\label{discussion}

In the present work we obtained statistical information about the matter power spectrum
by comparing the theoretical results for the
generalized Chaplygin gas model with the 2dFGRS observational data.
The free parameters of the model are
the equation of states parameters $\alpha$ and $\bar A$ and the
density parameters $\Omega_{dm0}$ and $\Omega_{c0}$. The complete
four-dimensional analysis is computationally hard but still
feasible. We have complemented it by a detailed study of the
cases for which only two or three parameters are free. This allows
us to verify the consistency and the correctness of the method
employed here.
\par
If the unification scenario with dark matter and dark energy as a
single fluid in a spatially flat universe is imposed from the
beginning (case (i)), the results of reference \cite{staro} are
essentially confirmed: there are parameter ranges for which the
data are well described by the generalized Chaplygin gas model.
The probability distribution function for $\alpha$ is high for
very small (near zero) or very large (greater than 2) values of
$\alpha$. Allowing the parameter $\bar A$ to vary, we find that
its one-dimensional PDF initially decreases with $\bar A$, but
increases as $\bar A = 1$ is approached. Notice that values
$\alpha > 1$ imply a superluminal sound speed and are therefore
unphysical (see, however, \cite{staro}).
\par
A different picture emerges for different priors. Leaving the
density parameters for the Chaplygin gas and the pressureless
matter components free, allows us to test the unified models
(pressureless matter is entirely baryonic) against models in which
there is separate dark matter, not accounted for by the Chaplygin
gas (cases (ii) and (iii)). We find that the unification scenario
is clearly disfavored. The PDF is highest in regions with very
small values for $\Omega_{c0}$ and large values for
$\Omega_{dm0}$. The behavior for $\alpha$ remains essentially the
same as in the previous case. This result  is confirmed when the
condition of a spatially flat universe is relaxed and both density
parameters are allowed to vary freely (case iv): the
one-dimensional PDFs for $\alpha$ and $\Omega_{c0}$ are decreasing
functions of $\alpha$ and $\Omega_{c0}$, respectively, while the
PDF for $\Omega_{dm0}$ increases with $\Omega_{dm0}$. Finally, the
full four-dimensional analysis of the phase space (case(v)) reproduces the
results for the lower dimensional cases (ii) - (iv).
\par
What is the origin of these apparently contradictory results? The first
aspect to be mentioned is that the matter power spectrum data only
poorly constrain the dark energy component. Even for the
$\Lambda$CDM model the matter power spectrum gives information
mainly on the dark matter component, the dark energy component
remaining largely imprecise. It is not by chance that the dark energy concept
emerged from the supernova data. Our results for
the Chaplygin gas model show  that a large amount of
dark matter, different from those described by the Chaplygin gas, is necessary to fit the data.
However, the
dispersion is quite high. For the flat case with a
three-dimensional parameter space we find at $2\sigma$, that
$\Omega_{dm0} = 1^{+0.00}_{-0.91}$.
Another point is the use of the neo-Newtonian formalism. However, for small values of the parameter $\alpha$, the main case of interest here, the differences to the full general relativistic treatment are not expected to be substantial. Moreover, in the cases of overlap the results of the full theory are reproduced.  Finally, possible statistical subtleties may influence the outcome of the investigation.
But as far as we could test the statistical
analysis (precision, crossing different information, etc), the
results seem to be robust. If this is really the case,
we must perhaps live with the fact that, while the SNe type Ia data favor  a unified model of the dark sector \cite{colistete}, this scenario is disfavored if large scale structure data are taken into account,
unless specific priors are imposed.

\vspace{1.0cm}

{\bf Acknowledgement}.  We thank FAPES and CNPq (Brazil) for
partial financial support (grants 093/2007 (CNPq and FAPES) and EDITAL
FAPES No. 001/2007). J.C.F. thanks also the french-brazilian
scientific cooperation CAPES/COFECUB and the Institut of
Astrophysique de Paris (IAP) for the kind hospitality during the
elaboration of this work.

\end{document}